\documentclass[11pt,prd,onecolumn,amsmath,amssymb,aps,floats,floatfix,nofootinbib]{revtex4-2}
\usepackage[colorlinks=true,urlcolor=blue,anchorcolor=blue,citecolor=blue,filecolor=blue,linkcolor=blue,menucolor=blue,linktocpage=true]{hyperref} 

\usepackage[inline]{enumitem}
\usepackage[multidot]{grffile}  
\usepackage{dcolumn}
\usepackage{bm}
\usepackage{amsmath}
\usepackage{amsfonts}
\usepackage{amssymb}
\usepackage{color}
\usepackage[table]{xcolor}
\usepackage{float}
\usepackage{latexsym}
\usepackage{slashed} 
\usepackage{pstricks}
\usepackage{indentfirst}
\usepackage{mathrsfs}
\usepackage{multirow}
\usepackage{epsfig,psfrag}
\usepackage{graphicx}
\usepackage{enumitem}
\usepackage{geometry,amssymb,yfonts}
\usepackage{yhmath}
\usepackage{anysize}
\usepackage{subfigure}
\usepackage{mathtools}
\usepackage{setspace} 
\usepackage[utf8]{inputenc} 
\DeclareUnicodeCharacter{FF3F}{_}
\usepackage[scientific-notation=true]{siunitx} 

\usepackage{url}
\usepackage{makecell}

\graphicspath{{fig/}}

\allowdisplaybreaks 

\newcommand{\Uone}{\mathrm{U}(1)}

\newcommand{\SUtwo}{\mathrm{SU}(2)}
\newcommand{\SO}{\mathrm{SO}}
\newcommand{\beq}{\begin{eqnarray}}
\newcommand{\eeq}{\end{eqnarray}}
\newcommand{\mcL}{\mathcal{L}}

\newcommand{\GeV}{\mathrm{GeV}}
\newcommand{\TeV}{\mathrm{TeV}}

\begin{document}

	\title{Leptophilic scalar dark matter in $\Uone_{L_\mu-L_\tau}$: Evading direct detection and prospective neutron star heating}

	\author{Chengfeng Cai$^{a}$}%
	\email[Contact author. ]{caichf3@mail.sysu.edu.cn}
	\author{Hong-Hao Zhang$^{b}$}
	\email[Contact author. ]{zhh98@mail.sysu.edu.cn}
	\affiliation{$^a$School of Science, Sun Yat-Sen University, Shenzhen 518107, China}
	\affiliation{$^b$School of Physics, Sun Yat-Sen University, Guangzhou 510275, China}

	\bigskip
	
	
	\begin{abstract}
	Leptophilic dark matter (DM) is a well-motivated thermal weakly interacting massive particle framework that can evade stringent nuclear-recoil searches while remaining testable via DM-induced heating of neutron stars (NSs). In this work, we study leptophilic scalar DM in a $\Uone_{L_\mu-L_\tau}$ gauge extension of the Standard Model, which provides a common leptophilic portal for all scenarios considered. To reproduce the observed relic abundance while suppressing direct-detection signals, we investigate three benchmark realizations: (i) a secluded DM scenario in which the relic density is set by annihilation into $\Uone_{L_\mu-L_\tau}$ gauge bosons and two pseudo-Nambu-Goldstone boson (pNGB) DM models based on (ii) an $\SO(4)$ symmetry and (iii) an $\SO(3)$ symmetry. In the
	$\SO(4)$ pNGB model, the DM mass arises at tree level from a soft breaking term, while the elastic scattering amplitude is suppressed by a symmetry-protected cancellation. In the $\SO(3)$ pNGB model, the DM mass is generated radiatively at one loop via the $\Uone_{L_\mu-L_\tau}$ gauge interaction, and we show that this gauging preserves the same cancellation mechanism, maintaining compatibility with direct-detection null results. We perform a systematic parameter scan imposing relic density, direct and indirect detection, and neutrino trident constraints and identify viable sub-TeV to TeV DM candidates. Under the optimistic maximal-heating assumption that the capture rate reaches the geometric limit and that the captured DM population efficiently thermalizes and attains capture-annihilation equilibrium inside NSs, we find that the remaining parameter space can be tested by near-infrared observations of old NSs, providing sensitivity complementary to terrestrial searches in regions that are currently weakly constrained.
	\end{abstract}

	\maketitle

	\section{Introduction}\label{sec:intro}
	The discovery of the Higgs boson at the Large Hadron Collider in 2012 marked the completion of the Standard Model (SM) particle spectrum \cite{ATLAS:2012yve,CMS:2012qbp}. Despite this triumph, the SM remains an incomplete description of nature, leaving fundamental questions unanswered, most notably the particle origin of dark matter (DM). Cosmological observations, particularly of the cosmic microwave background \cite{Planck:2018vyg}, establish that DM constitutes approximately $27\%$ of the Universe's energy budget. Yet, the SM provides no viable candidate. Unraveling the nature of DM and identifying its nongravitational interactions remain the most pressing challenges in high energy physics and cosmology.
	
	For decades, the weakly interacting massive particle (WIMP) has served as the leading paradigm, linking the observed relic abundance to the freeze-out mechanism governed by weak-scale couplings. However, this paradigm faces increasing tension with null results. A succession of direct detection experiments, including XENON~\cite{XENON:2023cxc}, LUX-ZEPLIN (LZ)~\cite{LZ:2024zvo}, PandaX~\cite{PandaX:2024qfu}, and CDEX~\cite{CDEX:2022kcd}, have yielded null results, imposing stringent constraints on the DM-nucleon scattering cross section. This creates a significant tension in simple WIMP models: the large couplings required for thermal freeze-out often imply scattering rates that are already ruled out. Consequently, reconciling the null results of direct detection with the requirement of a correct thermal relic density has become a central focus of DM phenomenology.
	
	To ameliorate this tension, theoretical efforts have followed two avenues. The first approach involves modifying the production mechanism itself, for instance, within the framework of ``secluded" dark matter. Here, the DM candidate $\chi$ couples feebly to the SM but interacts significantly with a secluded sector mediator $X$~\cite{Pospelov:2007mp}. The correct abundance is then achieved through annihilation channels such as $\chi\chi \to XX$, with variations including forbidden~\cite{DAgnolo:2015ujb}, not forbidden~\cite{Cline:2017tka}, and catalyzed annihilation~\cite{Xing:2021pkb, Cai:2021wmu} scenarios. The second approach focuses on suppressing the direct detection signal via symmetry arguments. A prominent example is the pseudo-Nambu-Goldstone boson (pNGB) DM model, where a softly broken $\Uone$ symmetry leads to a momentum-suppressed scattering amplitude, effectively canceling the signal at low velocities~\cite{Gross:2017dan}. This mechanism has been successfully extended to UV-complete U(1) theories~\cite{Abe:2020iph,Okada:2020zxo,Liu:2022evb}, non-Abelian symmetries~\cite{Alanne:2018zjm,Karamitros:2019ewv,Abe:2022mlc}, and vector-portal interactions~\cite{Cai:2021evx}.
	
	If terrestrial direct detection continues to report null results, astrophysical probes offer a complementary frontier. In particular, the capture of DM by compact stars, e.g., neutron stars, has emerged as a sensitive probe~\cite{Press:1985ug,Goldman:1989nd,Bertone:2007ae,Kouvaris:2007ay,Kouvaris:2010vv,McCullough:2010ai,deLavallaz:2010wp,McDermott:2011jp,Bell:2013xk,Bramante:2013nma,Bramante:2017xlb,Baryakhtar:2017dbj,Raj:2017wrv,Bell:2018pkk,Garani:2018kkd,Camargo:2019wou,Bell:2019pyc,Acevedo:2019agu,Joglekar:2019vzy,Joglekar:2020liw,Bell:2020jou,Ilie:2020vec,Maity:2021fxw,Bhattacharya:2023stq,Bhattacharya:2025xko}. The gravitational capture of DM transfers energy to the star, potentially heating old, cold neutron stars to temperatures ($\gtrsim 1000$ K), detectable by infrared observatories like the James Webb Space Telescope, TMT, or E-ELT~\cite{Baryakhtar:2017dbj,Raj:2024kjq}. Crucially, this method can be sensitive to interactions that standard direct detection misses. Since terrestrial experiments rely primarily on coherent scattering off nuclei, they are less sensitive to leptophilic DM. Neutron stars, however, contain a degenerate population of leptons (electrons and muons). 
	Therefore, a DM candidate that couples preferentially to leptons would naturally evade conventional direct detection bounds while remaining testable through the kinetic and annihilation heating of neutron stars.

The computational methodologies for evaluating DM capture in neutron stars have been systematically developed and refined in a series of works~\cite{Bell:2020jou,Bell:2020lmm,Anzuini:2021lnv}, including, in particular, a relativistic treatment that is important for lepton targets in neutron stars. Building on subsequent advances in neutron-star capture calculations, neutron-star heating has recently been explored as a probe of leptophilic dark matter in the context of a $\Uone_{L_\mu-L_\tau}$ gauge symmetry, one of the simplest anomaly-free extensions of the Standard Model~\cite{Foot:1990mn,He:1991qd}. An early study, Ref.~\cite{Garani:2019fpa}, investigated the capture of complex scalar dark matter in a secluded-sector framework. More recently, Ref.~\cite{Bell:2025acg} extended the analysis to fermionic dark matter and presented an updated capture calculation using the relativistic scattering formalism. Along this line, we extend the discussion of neutron star heating to leptophilic scalar dark matter and present three concrete realizations. One scenario follows the secluded paradigm as a warm-up and includes the relativistic scattering treatment. The other two are based on the pNGB DM framework. While the pNGB mechanism is well known to suppress nuclear-recoil signals via naturally realized destructive interference, its implementation in a genuinely leptophilic gauge sector has been explored less systematically, especially regarding the resulting neutron-star heating signals. We therefore perform a comprehensive study of these scalar scenarios, delineating the viable parameter space using the relic density requirement, direct- and indirect-detection limits~\cite{McDaniel:2023bju}, and constraints from neutrino trident production~\cite{Altmannshofer:2014pba}. Finally, we assess the optimistic maximal-heating reach of neutron-star observations for these models and show that near-infrared observations can probe leptophilic interactions that are currently weakly constrained by terrestrial experiments.

	The paper is organized as follows. In Section \ref{sec:model}, we provide a general discussion of neutron star capture rates via effective operators and introduce the specific model setups. Section \ref{sec:phenon} details the calculations for the relic density, indirect detection, direct detection, and neutrino trident constraints. In Section \ref{sec:results}, we present our numerical results and discuss the interplay between various experimental bounds and the neutron star heating signal. Finally, we summarize our findings in Section \ref{sec:summary}. 
	
\section{Leptophilic dark matter models}\label{sec:model}
\subsection{Effective operator analysis}\label{subsec:EFT}
To demonstrate the sensitivity of neutron stars (NSs) to dark matter candidates that evade conventional direct detection, we focus on a class of leptophilic scalar dark matter ($\chi$) models. Specifically, we consider scenarios where the effective interactions between DM and quarks, described by the dimension-5 ($\chi^\ast\chi \bar{q}q$) and dimension-6 $\left(\chi^{\ast} i \overleftrightarrow{\partial_\mu} \chi\right) \bar{q} \gamma^\mu q$ operators ($q=u,d$), are assumed to be negligible. Consequently, the stringent constraints from terrestrial direct detection experiments are naturally evaded. We assume that the communication between the dark sector and the SM is dominated by a dimension-6 current-current interaction, given by
\beq
\mcL_{\mathrm{dim6}}=\frac{1}{\Lambda^2} \left(\chi^{\ast} i \overleftrightarrow{\partial_\mu} \chi\right) \bar{l} \gamma^\mu l;\label{DMlepscattEff}
\eeq
where $\chi$ and $l$ represent the scalar dark matter and SM leptons, respectively. Through this interaction, DM particles can scatter off electrons or muons within compact objects such as neutron stars, leading to the capture and subsequent energy loss of the DM. Assuming that the kinetic and rest mass energy of the captured DM is fully thermalized and released via annihilation into SM particles, the NS surface temperature can be heated to over $2000$~K~\cite{Bell:2025acg}.

For a given equation of state (EoS) of the NS, the capture rate in the optically thin limit is given by~\cite{Bell:2020jou,Bell:2020lmm,Anzuini:2021lnv,Davoudiasl:2025gxn}
\beq
C & =&\frac{4 \pi}{v_{\star}} \frac{\rho_\chi}{m_{\mathrm{DM}}} \operatorname{Erf}\left(\sqrt{\frac{3}{2}} \frac{v_{\star}}{v_d}\right) \int_0^{R_{\star}} \sqrt{A(r)}\frac{\sqrt{1-B(r)}}{B(r)} \Omega^{-}(r) r^2 d r;
\eeq
where $A(r),~B(r)$ are defined by metric $ds^2=-B(r)dt^2+A(r)dr^2+r^2d\Omega^2$, and the interaction rate function $\Omega^{-}(r)$ is derived in Appendix~\ref{app:int_rate}:
\beq
	\Omega^-(r)=\frac{g_\ell\,\zeta_\ell(r)}{256\pi^3\,p_\chi(r)E_\chi(r)}	\int_{m_\ell}^{\mu_\ell(r)}dE_\ell \int_{s_-(E_\ell)}^{s_+(E_\ell)}ds	\int_{-\gamma^2(s)/s}^{0}dt	\frac{\overline{|\mathcal M|^2}(s,t)}{\gamma(s)}\overline{\mathcal P}_\ell(E_\ell,s,t;r).\label{scatrate}
\eeq 
Here, $g_\ell$ is the spin degeneracy factor, $\zeta_\ell(r)$ accounts for the target number density, and $\overline{\mathcal P}_\ell$ encodes Pauli blocking and the capture condition. 

Substituting the spin-averaged squared amplitude $|\overline{\mathcal{M}}|^2$ corresponding to the operator in Eq.~\eqref{DMlepscattEff},
\beq
|\overline{\mathcal{M}}|^2=\frac{4}{\Lambda^4}\left[\left(s-m_{\mathrm{DM}}^2-m_{\ell}^2\right)^2+t\left(s-m_{\ell}^2\right)\right];
\eeq
then we can compute the capture rate numerically. In this work, we adopt the BSk24 functional~\cite{Pearson:2018tkr} to describe the EoS of cold, nonrotating NSs, solving the Tolman-Oppenheimer-Volkoff equations~\cite{Tolman:1939jz,Oppenheimer:1939ne} with the central density $\rho_c$ as an input. The number density and chemical potential of muons are determined via $\beta$ equilibrium and charge conservation conditions (see Ref.~\cite{Bell:2020lmm} for details). Table~\ref{NSBenchmark} presents two benchmark NS configurations (BM-1 and BM-2) and the corresponding capture parameters derived in our analysis. Subsequent discussions on model sensitivities will be based on these representative NS profiles.

\begin{table}
	\centering
	\begin{tabular}{|c|c|c|c|c|c|}
		\hline
		& $\rho_c\left[\mathrm{g}\,\mathrm{cm}^{-3}\right]$
		& $M_{\star}\left[M_{\odot}\right]$
		& $R_{\star}\left[\mathrm{km}\right]$
		& $B\left(R_{\star}\right)$
		& $C\Lambda^4m_\chi~\left[\mathrm{GeV}^5\mathrm{s}^{-1}\right]$ \\
		\hline
		BM-1 & $7.76\times10^{14}$ & 1.5 & 12.593 & 0.648 & $2.76\times10^{39}$ \\
		\hline
		BM-2 & $1.04\times10^{15}$ & 1.9 & 12.415 & 0.548 & $1.65\times10^{40}$ \\
		\hline
	\end{tabular}
	\caption{Benchmark NS configurations derived using the BSk24 EoS. The last column lists the capture rate parameter independent of the specific cutoff scale and DM mass.}\label{NSBenchmark}
\end{table}

Numerical computation shows that for DM scattering off muon targets within the mass range $10~\GeV\lesssim m_{\mathrm{DM}}\lesssim 10~\TeV$, the capture rates are approximated by:
\beq
C \approx \left\{\begin{matrix}{2.76}\times10^{24}~\mathrm{s}^{-1}\left(\frac{\TeV}{\Lambda}\right)^4 \left(\frac{\TeV}{m_{\mathrm{DM}}}\right) & (\textrm{BM-1}:~M_\star=1.5~M_\odot)\\  {1.65}\times10^{25}~\mathrm{s}^{-1}\left(\frac{\TeV}{\Lambda}\right)^4 \left(\frac{\TeV}{m_{\mathrm{DM}}}\right) & (\textrm{BM-2}:~M_\star=1.9~M_\odot)\end{matrix}\right.\label{caprate}
\eeq
where $M_\star$ is the NS mass and $M_\odot$ is the solar mass.

While Eq.~\eqref{caprate} implies that the capture rate increases with interaction strength (smaller $\Lambda$), physically, the heating effect cannot grow indefinitely. The maximal heating of the neutron star is achieved when the capture process saturates to the so-called geometric limit. In this regime, the interaction is strong enough that every dark matter particle traversing the star is captured. The capture rate at this saturation point is determined solely by the geometrical cross section and the gravitational focusing of the NS, given by:
\beq
	C_{\mathrm{geom.}}=\frac{\pi R_\star^2\left[1-B\left(R_\star \right)\right]}{v_\star B\left(R_\star \right)} \frac{\rho_\chi}{m_{\mathrm{DM}}} \mathrm{Erf}\left(\sqrt{\frac{3}{2}} \frac{v_{\star}}{v_d}\right),\label{geolimit}
\eeq
where $\rho_\chi\approx0.4$ GeV/cm$^3$ is the local DM density, $v_\star=230$ km/s is the NS velocity, and $v_d=270$ km/s is the DM velocity dispersion. By equating the capture rate in Eq.~\eqref{caprate} to the geometric limit in Eq.~\eqref{geolimit}, we can derive the saturation threshold scale,{
\beq
\Lambda_{\ast} \approx \left\{\begin{matrix}2.95~\TeV&\qquad (\textrm{BM-1}:~M_\star=1.5~M_\odot)\\ 4.19~\TeV&\qquad (\textrm{BM-2}:~M_\star=1.9~M_\odot)\end{matrix}\right..
\eeq
This energy scale is within the reach of future colliders, such as the prospective muon collider~\cite{Huang:2021nkl}. If the operator in Eq.~\eqref{DMlepscattEff} arises from the exchange of a BSM gauge boson $X_\mu$, this translates to an optimistic sensitivity in the parameter space of $m_X/g_X\sim \Lambda_\ast$. 

We will use $\Lambda_\ast$ as the characteristic scale entering the optimistic maximal-heating benchmark adopted in the phenomenological analysis below. This benchmark assumes efficient energy deposition by the captured DM population and capture-annihilation equilibrium inside the NS. For capture-saturating interactions, previous studies have found that kinetic energy deposition is typically rapid and that capture-annihilation equilibrium is reached well within the age of an old NS for broad classes of $s$-wave annihilating DM models~\cite{Bell:2023ysh}. }

Within the EFT framework, operators such as $\chi^\ast\chi \bar{q}q$ and $\left(\chi^{\ast} i \overleftrightarrow{\partial_\mu} \chi\right) \bar{q} \gamma^\mu q$ can be manually set to zero. However, realizing such suppression in a concrete model is nontrivial. Suppressing the vector current operator $\left(\chi^{\ast} i \overleftrightarrow{\partial_\mu} \chi\right) \bar{q} \gamma^\mu q$ is relatively straightforward; if the interaction is mediated by a leptophilic gauge field [e.g., $\Uone_{L_\mu-L_\tau}$], the quark coupling arises only via kinetic mixing at the loop level and is naturally suppressed. Conversely, suppressing the scalar operator $\chi^\ast\chi \bar{q}q$ is more subtle, as it is typically generated via the Higgs portal. In many simple scenarios, the Higgs portal is essential for DM annihilation ($\chi^\ast\chi\to h_{i}\to \text{SM}$ or $\chi^\ast\chi\to h_i h_j$) to satisfy relic density constraints. Thus, a tension exists between avoiding direct detection bounds (which requires a small Higgs portal coupling) and achieving the correct relic abundance.

In the following subsections, we introduce three specific model realizations that resolve this tension:
\begin{enumerate}
	\item Secluded dark matter~\cite{Pospelov:2007mp,Su:2025mxv}: We consider a complex scalar DM, $\chi$, charged under a $\Uone_{L_\mu-L_\tau}$ gauge symmetry with the mass relation $m_{DM}>m_X$. Here, the DM relic density is determined by the annihilation channel $\chi^\ast\chi\to XX$ within the dark sector, allowing the Higgs portal coupling to be negligible. The gauge boson $X_\mu$ subsequently decays into SM particles via kinetic mixing. This setup reproduces the observed relic density via the standard freeze-out mechanism while evading nuclear scattering constraints, leaving NS heating as a key probe. A concrete realization is presented in Subsection~\ref{subsect:SSDM}.
	\item Pseudo-Nambu-Goldstone boson dark matter: We assume a complex scalar DM $\chi$, charged under $\Uone_{L_\mu-L_\tau}$, is embedded in a multiplet of a spontaneously broken global symmetry [e.g., $\SO(N)$], identifying $\chi$ as a Goldstone boson. Explicit breaking of this global symmetry generates a mass for $\chi$. Due to the approximate shift symmetry of the pNGB, the $\chi^\ast\chi \bar{q}q$ effective operator is naturally suppressed, leading to a cancellation in the $\chi$-nucleon scattering amplitude at low momentum transfer~\cite{Gross:2017dan}.\footnote{Note that NS heating for pNGB DM was previously explored in the context of the conventional Higgs portal~\cite{Zeng:2021moz}. In such setups, the DM-nucleon scattering responsible for NS capture suffers from the same momentum-dependent suppression as in direct detection, limiting its primary advantage to the low-mass regime. In contrast, our leptophilic framework disentangles the capture dynamics: The tree-level DM-muon scattering mediated by $X_\mu$ is completely unaffected by the pNGB cancellation mechanism. This unsuppressed capture process allows the NS heating sensitivity to surpass direct detection limits even in the sub-TeV and TeV mass ranges.} We discuss two distinct realizations of this scheme: in Subsection~\ref{subsect:pNGBA}, we present a model based on an $\SO(4)$ global symmetry where the DM mass is generated at the tree level via soft breaking terms; in Subsection~\ref{subsect:pNGBB}, we explore an $\SO(3)$ symmetric framework where the DM mass arises radiatively at the loop level.  
\end{enumerate}

\subsection{Secluded scalar dark matter model}\label{subsect:SSDM}
Following a framework similar to the one presented in Refs.~\cite{Garani:2019fpa,Su:2025mxv}, we introduce a complex scalar dark matter candidate, $\chi$, which carries a $\Uone_{L_\mu-L_\tau}$ gauge charge $Q_{DM}$. The relevant Lagrangian for the dark sector is given by
\beq
\mathcal{L}=  -\frac{1}{4} X^{\mu \nu} X_{\mu \nu}+\frac{1}{2} m_{X}^2 X^{ \mu} X_\mu+ |D_\mu \chi|^2 -m_{DM}^2|\chi|^2-\lambda_{\chi}|\chi|^4-\lambda_{\chi H}|\chi|^2|H|^2,\eeq
where the covariant derivative is defined as:
\beq
D_\mu\chi=\left(\partial_\mu-iQ_{DM}g_XX_\mu\right)\chi.
\eeq

For simplicity, we assume the mass of the gauge boson $X_\mu$ is generated via the Stückelberg mechanism~\cite{Stueckelberg:1938hvi,Stueckelberg:1938zz,Ruegg:2003ps}, avoiding the need to introduce additional scalar degrees of freedom for symmetry breaking. In this setup, by setting the Higgs portal coupling $\lambda_{\chi H}$ to be negligibly small, the tree-level DM-nucleon scattering cross section is highly suppressed. Consequently, the dominant contribution to nuclear scattering arises purely from the kinetic mixing between $X_\mu$ and the SM gauge bosons. As we will discuss in detail in Section~\ref{subsect:DD}, this mixing is loop induced in the $\Uone_{L_\mu-L_\tau}$ model and is, therefore, naturally small, safely evading current direct detection bounds.

In the secluded mass regime, where the DM is heavier than the mediator ($m_{DM} > m_X$), the relic abundance is primarily determined by the annihilation process $\chi^\ast \chi \to X X$ into on-shell gauge bosons. While DM pairs can also annihilate into SM leptons ($\ell = \mu, \tau$ and their corresponding neutrinos) via $s$-channel $X_\mu$ exchange ($\chi^\ast \chi \to \bar{\ell}\ell, \bar{\nu}\nu$), these processes are $p$-wave suppressed for a scalar DM candidate. Because the $p$-wave annihilation cross section scales with the velocity squared ($v^2 \sim x^{-1}$, where $x = m_{DM}/T$), these channels remain subdominant throughout most of the parameter space during freeze-out. In Section \ref{sec:results}, we will present the detailed formulas for these annihilation cross sections.

At low energies, integrating out the heavy gauge field $X_\mu$ yields the dimension-6 effective operator, Eq.~\eqref{DMlepscattEff}, connecting the dark matter and lepton currents. By matching the amplitudes of the full and effective theories, the characteristic cutoff scale $\Lambda$ is identified as
\beq
\Lambda=\frac{m_X}{\sqrt{Q_{DM}}g_X}.\label{effLambda1}
\eeq
	
\subsection{pNGB dark matter model A}\label{subsect:pNGBA}
In this model, we consider a pNGB DM paradigm~\cite{Gross:2017dan} featuring a soft symmetry-breaking mass term. The scalar sector is extended by a dark doublet $\Phi=(\Phi_1,\Phi_2)^T=(1/\sqrt{2})(\phi_1-i\phi_2,\phi_3-i\phi_4)^T$. We first consider the tree-level scalar potential. By analogy with the custodial symmetry of the SM Higgs, the potential of $\Phi$ can be constructed to accidentally respect a global $\SUtwo\times \SUtwo\simeq \SO(4)$ symmetry. The tree-level potential is assumed to be~\cite{Abe:2022mlc}:
\beq V_{\text {tree }}^{(\mathrm{IR})}\left(H, \Phi\right)&=&-\mu_\phi^2\Phi^\dag \Phi+\lambda_\phi\left(\Phi^\dag \Phi\right)^2-\mu_H^2\left(H^{\dagger} H\right)+\lambda_H\left(H^{\dagger} H\right)^2 \nonumber\\
&&+2 \lambda_{H \phi}\left(H^{\dagger} H\right)\left(\Phi^\dag \Phi\right) +m^2\Phi^\dag\sigma^3 \Phi \nonumber\\
&=& -\mu_\phi^2\left(\left|\Phi_1\right|^2+\left|\Phi_2\right|^2\right)+\lambda_\phi\left(\left|\Phi_1\right|^2+\left|\Phi_2\right|^2\right)^2 \nonumber\\
&&-\mu_H^2\left(H^{\dagger} H\right)+\lambda_H\left(H^{\dagger} H\right)^2+2 \lambda_{H \phi}\left(H^{\dagger} H\right)\left(\left|\Phi_1\right|^2+\left|\Phi_2\right|^2\right) \nonumber\\
&&+m^2\left(\left|\Phi_1\right|^2-\left|\Phi_2\right|^2\right)~,\label{VSO4IR}
\eeq
where the first two lines manifestly preserve the $\SO(4)$ symmetry, and the final $m^2$ term represents a soft breaking of the symmetry. 

We assume the potential triggers spontaneous symmetry breaking such that $\langle\phi_3\rangle=v_\phi$ and $\langle H \rangle = v$. Expanding the fields around their vacuum expectation values (VEVs),
\beq
\phi_3=v_\phi+s, \quad H=\begin{pmatrix}H^+\\ \frac{v+h+iG}{\sqrt{2}}\end{pmatrix},
\eeq
the $\SO(4)$ symmetry is spontaneously broken to $\SO(3)$, generating Goldstone bosons. We identify the complex scalar $\Phi_1 \equiv \chi$ as the DM candidate.

From the minimization condition with respect to $\Phi_2$ ($\partial V / \partial \Phi_2 = 0$), we obtain the relation $\mu_\phi^2 - \lambda_\phi v_\phi^2 - {\lambda_{H\phi}v^2 }= -m^2$. Substituting this back into the quadratic terms for $\Phi_1$ in Eq.~\eqref{VSO4IR}, the DM mass is generated entirely by the soft-breaking parameter,
\beq
m_{\mathrm{DM}}^2=2m^2.
\eeq
Since the $\SO(4)$ symmetry is broken only softly in the potential, the approximate shift symmetry of the pNGB is preserved, ensuring the cancellation mechanism for the DM-nucleon scattering amplitude at zero momentum transfer~\cite{Abe:2022mlc}.

To couple the dark sector to the SM leptons and realize the NS heating signal, we gauge a $\Uone_{L_\mu-L_\tau}$ subgroup. The $\Uone_{L_\mu-L_\tau}$ transformations of the two components of $\Phi$ are assigned as follows:
\beq
\begin{pmatrix}\Phi_1\\ \Phi_2\end{pmatrix}\to \begin{pmatrix}e^{iQ_{DM}\theta}&0\\ 0&e^{iQ_2\theta}\end{pmatrix} \begin{pmatrix}\Phi_1\\ \Phi_2\end{pmatrix}~,
\eeq
leading to the covariant derivative
\beq
D_\mu\Phi=\partial_\mu\Phi-ig_X X_\mu \begin{pmatrix}Q_{DM}&0\\0&Q_2\end{pmatrix}\Phi~.
\eeq
Once $\phi_3$ acquires its VEV $v_\phi$, the $\Uone_{L_\mu-L_\tau}$ gauge symmetry is spontaneously broken, and the component $\phi_4$ is eaten by the gauge boson $X_\mu$, giving it a mass $m_X = |Q_2| g_X v_\phi$. Importantly, any asymmetry in the gauge charges ($|Q_{DM}| \neq |Q_2|$) would induce different radiative corrections to the Higgs portal couplings of $\Phi_1$ and $\Phi_2$, thereby ruining the delicate cancellation in the DM-nucleon scattering amplitude. We therefore mandate the specific charge assignment $Q_2=\pm Q_{DM}$. 

Note that this entire low-energy architecture—both the restricted form of the scalar potential and the specific charge assignment—is not an ad hoc construction but can naturally emerge from an ultraviolet (UV) completion. Suppose the dark sector originates from a gauged $\SUtwo_X$ symmetry at a high scale $\Lambda_{UV}$, where $\Phi$ transforms as a fundamental doublet. First, just like the SM Higgs, the renormalizable potential of an $\SUtwo_X$ doublet is intrinsically restricted to an $\SO(4)$-invariant form; the global $\SO(4)$ symmetry is accidental and not imposed by hand. Second, the two complex components of an $\SUtwo_X$ doublet naturally carry opposite eigenvalues under its diagonal generator ($T^3$). Thus, when $\SUtwo_X$ is broken down to the $\Uone_{L_\mu-L_\tau}$ subgroup, the relation $Q_{DM} = -Q_2$ is automatically enforced. Finally, this symmetry breaking can be triggered by an $\SUtwo_X$ triplet scalar $\Sigma^a$ acquiring a VEV $\langle \Sigma^a\rangle=v_\Sigma\delta^{a3}$. The interaction $\kappa\Phi^\dag \sigma^a\Phi \Sigma^a$ then generates the requisite soft-breaking mass term $m^2 \propto \kappa v_\Sigma$ in the infrared (IR). Constructing the full UV theory is beyond the scope of this work. Our analysis is, therefore, restricted to the DM phenomenology, which requires only the assumed low-energy field content and symmetries and is otherwise independent of UV details.

The CP-even scalars $(h, s)$ mix to form the physical mass eigenstates $(h_1, h_2)$. The mass matrix is given by
\beq
&&\mathcal{L} \supset-\frac{1}{2}\begin{pmatrix}
	h & s\end{pmatrix}M_h^2\binom{h}{s}=-\frac{1}{2}\begin{pmatrix}
	h & s\end{pmatrix}\begin{pmatrix}
	m_h^2 & m_{h s}^2 \\
	m_{h s}^2 & m_s^2
\end{pmatrix}\binom{h}{s}, \\
&&m_h^2=2 \lambda_H v^2, \quad m_s^2=2 \lambda_\phi v_\phi^2, \quad m_{h s}^2=2 \lambda_{H \phi} v v_\phi,\nonumber
\eeq
which is diagonalized by an orthogonal matrix $O$,
\beq
&&{ M_h^{\prime 2}=O^T M_h^2 O=\begin{pmatrix}
	m_{h_1}^2 & 0 \\
	0 & m_{h_2}^2
\end{pmatrix} } ,\quad \binom{h_1}{h_2}=O^T\binom{h}{s},\\
&&O=\begin{pmatrix}
	c_\theta & s_\theta \\
	-s_\theta & c_\theta
\end{pmatrix}\qquad t_{2 \theta}=\frac{2 m_{h s}^2}{m_{s}^2-m_h^2}
\eeq
where $c_\theta\equiv \cos\theta$, $s_\theta\equiv \sin\theta$, and $t_{2\theta}\equiv \tan 2\theta$.

For the phenomenological discussions in Section~\ref{sec:phenon}, we summarize the relevant couplings in the mass basis. The scalar self-interactions involving the DM candidate $\chi$ and the mediators $h_{1,2}$ are
\beq
\mathcal{L} &\supset&-g_{\chi \chi h_1} h_1|\chi|^2-g_{\chi \chi h_2} h_2|\chi|^2\nonumber\\
&&-\frac{1}{2} g_{\chi \chi h_1 h_1} h_1^2|\chi|^2-\frac{1}{2} g_{\chi \chi h_2 h_2} h_2^2|\chi|^2-g_{\chi \chi h_1 h_2} h_1 h_2|\chi|^2\nonumber\\
&&-\frac{1}{3!} g_{h_1 h_1 h_1} h_1^3-\frac{1}{2!} g_{h_1 h_1 h_2} h_1^2 h_2-\frac{1}{2!} g_{h_1 h_2 h_2} h_1 h_2^2-\frac{1}{3!} g_{h_2 h_2 h_2} h_2^3
\eeq
where the effective couplings are defined as
\beq
&& g_{\chi\chi h_1}=2 \left(\lambda_{H\phi} v c_\theta-\lambda_{\phi} v_\phi s_\theta\right)~,\qquad g_{\chi\chi h_2}=2 \left(\lambda_{H\phi} v s_\theta+\lambda_{\phi} v_\phi c_\theta\right),\\
&& g_{\chi\chi h_1 h_1}=2 \left(\lambda_{H \phi} c_\theta^2+\lambda_{\phi} s_\theta^2\right),\qquad g_{\chi\chi h_2 h_2}=2 \left(\lambda_{H \phi} s_\theta^2+\lambda_{\phi} c_\theta^2\right),\nonumber\\
&& g_{\chi\chi h_1 h_2}=2 s_\theta c_\theta\left(\lambda_{H \phi}-\lambda_{\phi}\right),\\
&& g_{h_1 h_1 h_1}  =6\left[\lambda_H v c_\theta^3-\lambda_{\phi} v_\phi s_\theta^3+\lambda_{H \phi}\left(v_\phi c_\theta^2 s_\theta-v c_\theta s_\theta^2\right)\right], \nonumber\\
&& g_{h_2 h_2 h_2}  =6\left[\lambda_H v s_\theta^3+\lambda_{\phi} v_\phi c_\theta^3+\lambda_{H \phi}\left(v_\phi s_\theta^2 c_\theta+v s_\theta c_\theta^2\right)\right], \nonumber\\
&& g_{h_1 h_1 h_2}  =6 \lambda_H v c_\theta^2 s_\theta+6 \lambda_{\phi} v_\phi s_\theta^2 c_\theta+\lambda_{H \phi}\left[2 v_\phi\left(c_\theta^3-2 c_\theta s_\theta^2\right)+2 v\left(s_\theta^3-2 c_\theta^2 s_\theta\right)\right], \nonumber\\
&& g_{h_1 h_2 h_2} =6 \lambda_H v c_\theta s_\theta^2-6 \lambda_{\phi} v_\phi s_\theta c_\theta^2+\lambda_{H \phi}\left[2 v_\phi\left(s_\theta^3-2 c_\theta^2 s_\theta\right)-2 v\left(c_\theta^3-2 c_\theta s_\theta^2\right)\right].
\eeq

Finally, the interactions between the scalar mediators $h_i$ and the SM fermions ($f=t,b, \dots$) or gauge bosons ($V=W^\pm, Z$) inherit from the SM Higgs component via the mixing angle $\theta$,
\beq
\mathcal{L}_{\mathrm{Yuk.,i}}&\supset& \sum_f g_{h_iff}h_i\bar{f}f,\quad g_{h_1ff}=-\frac{m_f}{v}c_\theta,\quad g_{h_2ff}=-\frac{m_f}{v}s_\theta,\label{yukcoup}\\
\mathcal{L}_{\mathrm{Gauge}}&\supset& g_{h_iWW}h_i W^+_\mu W^{-,\mu}+\frac{1}{2}g_{h_iZZ}h_i Z_\mu Z^\mu,\quad g_{h_1VV}=\frac{2m_V^2}{v}c_\theta,\quad g_{h_2VV}=\frac{2m_V^2}{v}s_\theta.\label{gaugecoup}
\eeq

\subsection{pNGB dark matter model B}\label{subsect:pNGBB}
In the second benchmark model, we consider a real scalar triplet $\Phi^a$ ($a=1,2,3$) with a global $\SO(3)$ symmetry at the UV matching scale $\Lambda_{UV}$. Below $\Lambda_{UV}$, the $\SO(3)$ symmetry is explicitly broken only by the gauging of the $\SO(2)_{L_\mu-L_\tau}\simeq\Uone_{L_\mu-L_\tau}\subset \SO(3)$ subgroup, which corresponds to rotations in the $(\Phi^1,\Phi^2)$ components. We assume the $\Uone_{L_\mu-L_\tau}$ gauge boson $X_\mu$ obtains a Stückelberg mass $m_X$. In the linear representation, we define the complex field
\beq
\chi \equiv \frac{\Phi^1+i \Phi^2}{\sqrt{2}}, \quad {\hat{s} \equiv \Phi^3}, \quad D_\mu \chi=\left(\partial_\mu-i Q_{DM} g_X X_\mu\right) \chi.
\eeq
In the $\SO(3)$ limit, the renormalizable scalar potential can be written as
\beq
V_{\mathrm{UV}}\left(H, \Phi^a\right)=m_H^2 H^{\dag} H+\lambda_H\left(H^{\dag} H\right)^2+\frac{m_{\Phi}^2}{2} \Phi^a \Phi^a+\frac{\lambda_{\Phi}}{4}\left(\Phi^a \Phi^a\right)^2+\lambda_{H \Phi}\left(H^{\dag} H\right) \Phi^a \Phi^a,\label{eq:SO3_def_Phi_landau}
\eeq
where $m_\Phi^2<0$ triggers the spontaneous symmetry-breaking $\SO(3)\to \Uone_{L_\mu-L_\tau}$.
We choose the aligned vacuum
\beq
\langle H\rangle=\frac{1}{\sqrt{2}}\binom{0}{v}, \quad\langle {\hat{s} }\rangle=v_\phi, \quad\langle\chi\rangle=0, \quad H=\frac{1}{\sqrt{2}}\binom{0}{v+h}, \quad {\hat{s} =v_\phi+s},\label{eq:vev_param_landau_new}
\eeq
so that $\chi$ remains a massless Goldstone boson at the tree level.

Below $\Lambda_{UV}$, once $\Uone_{L_\mu-L_\tau}$ is gauged, the $\SO(3)$ symmetry is explicitly broken. At the loop level, gauge interactions modify the potential and the field strength of $\chi$, inducing $\SO(3)$-breaking terms. The most important consequence is that a mass for the pNGB $\chi$ is radiatively generated. This mass can be computed by integrating out the gauge field $X_\mu$ and adding the one-loop Coleman-Weinberg potential originating from the gauge loop:
\beq
V_{1-\mathrm{loop}}^{(X)}=\frac{3}{64 \pi^2} M_X^4(\chi)\left[\ln \frac{M_X^2(\chi)}{\mu^2}-\frac{5}{6}\right]+V_{\mathrm{CT}},
\eeq
where $V_{\mathrm{CT}}$ is a counterterm fixing the UV behavior. The field-dependent mass of the gauge field is given by
\beq
M_X^2(\chi)=m_X^2+2\left(Q_{DM} g_X\right)^2 \chi^{\ast} \chi.
\eeq
Expanding $V_{1-\mathrm{loop}}^{(X)}$, we can find a correction to the quadratic term of $\chi$: 
\beq
-\mathcal{L}\supset \delta m_\chi^2 \chi^{\ast} \chi.
\eeq
Applying the matching condition
\beq
\delta m_\chi^2(\Lambda_{UV})=0
\eeq
at $\Lambda_{UV}$ where the $\SO(3)$ symmetry is restored, we find the mass of the DM candidate $\chi$ to be
\beq
m_{DM}^2\equiv \delta m_\chi^2(m_X)\approx\frac{3\left(Q_{DM} g_X\right)^2}{16 \pi^2} m_X^2 \ln \left(\frac{\Lambda_{UV}^2}{m_X^2}\right)\label{DMmass_loop}
\eeq
at an IR scale below $m_X$. Note that the field strength renormalization of $\chi$ would rescale $m_{DM}^2$ by a factor $Z_\chi^{-1}\simeq [1+\mathcal{O}(g_X^2/16\pi^2)]$. Since this is a next-to-leading logarithmic effect, it can be neglected for the mass term calculation.

In contrast to model A, the symmetry-breaking mass term here is generated radiatively due to gauge interactions. Because our motivation for using a pNGB DM model is to automatically suppress the DM-quark scattering amplitude in the zero-momentum-transfer limit, we must verify that these loop corrections do not spoil the cancellation mechanism. In the IR effective theory, the most general renormalizable potential consistent with $\Uone_{L_\mu-L_\tau}$ [but not necessarily $\SO(3)$] is parametrized as {
\beq
V_{\mathrm{IR}}(H,\hat{s},\chi)&=&m_H^2H^\dag H+\lambda_H(H^\dag H)^2+\frac{m_3^2}{2}\hat{s}^2+\frac{\lambda_{33}}{4}\hat{s}^4
+m_\chi^2\chi^{\ast} \chi+\lambda_\chi(\chi^{\ast} \chi)^2 \nonumber\\
&&+\lambda_{H3}\hat{s}^2 H^\dag H+2\lambda_{H\chi}(H^\dag H)(\chi^{\ast} \chi)
+\lambda_{3\chi}\hat{s}^2(\chi^{\ast} \chi),\label{eq:V_linear_RGE_landau_main}
\eeq }
and the following matching conditions are imposed at $\Lambda_{UV}$ to recover the $\SO(3)$ limit:
\beq
&&m_\chi^2(\Lambda_{UV})=m_3^2(\Lambda_{UV})=m_\Phi^2\Rightarrow \delta m_\chi^2(\Lambda_{UV})\equiv m_\chi^2(\Lambda_{UV})-m_3^2(\Lambda_{UV})=0,\nonumber\\
&&\lambda_{33}(\Lambda_{UV})=\lambda_\chi(\Lambda_{UV})=\lambda_{3\chi}(\Lambda_{UV})=\lambda_\Phi,\qquad\lambda_{H3}(\Lambda_{UV})=\lambda_{H\chi}(\Lambda_{UV})=\lambda_{H\Phi},
\label{eq:SO3_UV_BC_landau_main}
\eeq

\begin{figure}
	\centering
	\includegraphics[width=0.9\textwidth]{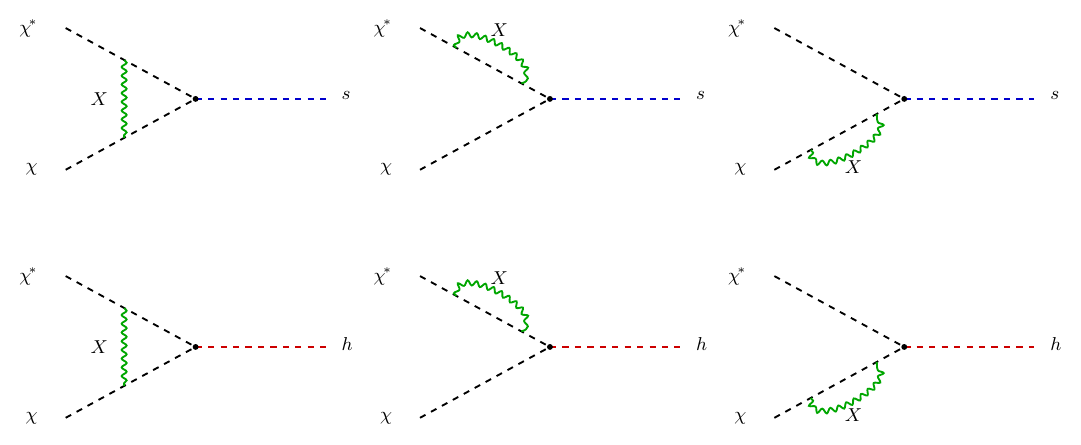}
	\caption{Loop corrections to the trilinear Higgs portal couplings.}\label{feyndiag_modelB_1}
\end{figure}

The masses of the Higgs-portal mediators, $m_{h}^2$, $m_s^2$, and their mixing, $m_{hs}^2$, are not corrected by gauge interactions at the one-loop level, since $h$ and $s$ are neutral under $\Uone_{L_\mu-L_\tau}$. On the other hand, { the $\chi$-dependent Higgs-portal couplings $\lambda_{3\chi}$ and $\lambda_{H\chi}$, which generate the $\chi^\ast\chi s$ and $\chi^\ast\chi h$ vertices after symmetry breaking,} are renormalized via the $\chi$ self-energy and vertex corrections involving the gauge boson. The most relevant Feynman diagrams for these corrections are shown in Fig.~\ref{feyndiag_modelB_1}. The gauge loops universally scale these vertices, meaning $\lambda_{3\chi}$ and $\lambda_{H\chi}$ are rescaled by the same factor $r$ compared to $\lambda_{33}$ and $\lambda_{H3}$. This factor is given by
\beq
\lambda_{3\chi}^{(X)}\approx r\lambda_{33},\qquad \lambda_{H\chi}^{(X)}\approx r\lambda_{H3},\qquad r \approx \frac{1}{1-\frac{3\left(Q_{DM} g_X\right)^2}{16 \pi^2} \ln \frac{\Lambda_{UV}^2}{m_X^2}}\approx\frac{1}{1-\frac{m_{DM}^2}{m_X^2}}\label{rescale},
\eeq
where the superscript $(X)$ denotes quantities corrected by loops involving the $X_\mu$ gauge boson.

Computing the DM-quark scattering amplitude in the Mandelstam $t\to0$ limit, we find
\beq
\mathcal{M}(0)\propto \left(g_{\chi \chi h},~ g_{\chi \chi s}\right)\left(-\mathbf{M}^2\right)^{-1}\binom{m_q / v}{0},
\eeq
where $g_{\chi \chi h}=2\lambda_{H\chi}^{(X)}v$, $g_{\chi \chi s}=2\lambda_{3\chi}^{(X)}v_\phi$, and
\beq
\mathbf{M}^2=\begin{pmatrix}
	2\lambda_H v^2 & 2 \lambda_{H 3} v v_\phi \\
	2 \lambda_{H 3} v v_\phi & 2 \lambda_{33} v_\phi^2
\end{pmatrix}
\eeq
is the mass matrix of the CP-even scalar fields $(h,s)$. Since $\lambda_{H\chi}=\lambda_{H3}=\lambda_{H\Phi}$ and $\lambda_{3\chi}=\lambda_{33}=\lambda_\Phi$ at the tree level, we obtain
\beq
\mathcal{M}(0)\propto \frac{ m_q }{\left(\lambda_H \lambda_{33}-\lambda_{H3}^2\right)v^2} \left(\lambda_{3\chi}^{(X)}\lambda_{H3}-\lambda_{33}\lambda_{H\chi}^{(X)}\right)\to0\label{amplitude_t_0}
\eeq
at the leading logarithmic order. This demonstrates that the cancellation mechanism remains effective in this model. The validity of the cancellation relies on the relation in Eq.~\eqref{rescale}. In Appendix~\ref{RGEs}, we provide the renormalization group equations (RGEs) for the couplings, confirming Eq.~\eqref{amplitude_t_0} through both numerical solutions and an approximate analytical analysis.

The scalar self-couplings are analogous to those in model A, with the corresponding $\lambda_i$ subscripts adapted for this model and the universal rescaling factor $r$ applied to all vertices involving $\chi$. Specifically,
\beq
&& g_{\chi\chi h_1}=2 \left(\lambda_{H3} v c_\theta-\lambda_{33} v_\phi s_\theta\right)r,\qquad g_{\chi\chi h_2}=2 \left(\lambda_{H3} v s_\theta+\lambda_{33} v_\phi c_\theta\right) r\\
&& g_{\chi\chi h_1 h_1}=2 \left(\lambda_{H 3} c_\theta^2+\lambda_{33} s_\theta^2\right)r,\qquad g_{\chi\chi h_2 h_2}=2 \left(\lambda_{H 3} s_\theta^2+\lambda_{33} c_\theta^2\right)r,\nonumber\\
&& g_{\chi\chi h_1 h_2}=2 s_\theta c_\theta\left(\lambda_{H 3}-\lambda_{33}\right)r,\\
&& g_{h_1 h_1 h_1}=6\left[\lambda_H v c_\theta^3-\lambda_{33} v_\phi s_\theta^3+\lambda_{H 3}\left(v_\phi c_\theta^2 s_\theta-v c_\theta s_\theta^2\right)\right], \nonumber\\
&& g_{h_2 h_2 h_2}=6\left[\lambda_H v s_\theta^3+\lambda_{33} v_\phi c_\theta^3+\lambda_{H 3}\left(v_\phi s_\theta^2 c_\theta+v s_\theta c_\theta^2\right)\right], \nonumber\\
&& g_{h_1 h_1 h_2}=6 \lambda_H v c_\theta^2 s_\theta+6 \lambda_{33} v_\phi s_\theta^2 c_\theta+\lambda_{H 3}\left[2 v_\phi\left(c_\theta^3-2 c_\theta s_\theta^2\right)+2 v\left(s_\theta^3-2 c_\theta^2 s_\theta\right)\right], \nonumber\\
&& g_{h_1 h_2 h_2}=6 \lambda_H v c_\theta s_\theta^2-6 \lambda_{33} v_\phi s_\theta c_\theta^2+\lambda_{H 3}\left[2 v_\phi\left(s_\theta^3-2 c_\theta^2 s_\theta\right)-2 v\left(c_\theta^3-2 c_\theta s_\theta^2\right)\right].
\eeq

On the other hand, the CP-even scalars couple to the SM fermions and gauge fields in the same manner as in the previous model, with the corresponding couplings given by Eqs.~\eqref{yukcoup} and \eqref{gaugecoup}.

\section{Phenomenology}\label{sec:phenon}
\subsection{Relic density and indirect detection}
The thermal relic abundance of the dark matter is determined by its annihilation cross sections in the early Universe. We first outline the dominant annihilation channels for each model under consideration.

For the secluded DM model, assuming $m_{\mathrm{DM}} > m_X$, the primary annihilation process occurs entirely within the dark sector:
\begin{enumerate}
	\item $\chi^\ast+\chi\to X+X$. According to Ref.~\cite{Su:2025mxv}, the velocity-averaged annihilation cross section is given by {
	\beq
	\left\langle\sigma v\right\rangle_{XX}=\frac{Q_{\mathrm{DM}}^4g_X^4\left(1-\xi_X^{2}+\frac{3}{8}\xi_X^{4}\right)}{8 \pi m_{\mathrm{DM}}^2\left(1- \frac{1}{2}\xi_X^2\right)^2} \sqrt{1-\xi_X^2},
	\eeq }
	where we define the mass ratio $\xi_X \equiv m_X/m_{\mathrm{DM}}$.
	
	\item $\chi^\ast+\chi\to X^\ast \to \bar{\ell}+\ell$. Annihilation into SM leptons ($\ell=\mu,\tau,\nu_{\mu},\nu_{\tau}$) via $s$-channel $X_\mu$ exchange yields:{
	\beq
	\langle\sigma_{\bar{\ell}\ell} v\rangle\approx\sum_{\ell}\frac{1}{\pi n_\ell}\frac{Q_{\mathrm{DM}}^2 g_X^4}{(4m_{\mathrm{DM}}^2-m_{X}^2)^2+m_X^2\Gamma_X^2}m_{\mathrm{DM}}^2\sqrt{1-\xi_\ell^2}\left(1+\frac{1}{2}\xi_\ell^2\right)x^{-1},
	\eeq}
	where $n_\ell=1(2)$ for $\ell=\mu,\tau$ ($\nu_{\mu,\tau}$), and $\xi_\ell \equiv m_\ell/m_{\mathrm{DM}}$. Because $\chi$ is a scalar, this process is $p$ wave and, therefore, velocity suppressed by a factor of $x^{-1} \equiv T/m_{\mathrm{DM}}$. Consequently, these channels become significant only when $m_{X}\lesssim \mathcal{O}(100 \text{ GeV})$ and $m_{\mathrm{DM}} < m_W$. In the parameter space of interest for this work, this contribution is generally subdominant during freeze-out.
\end{enumerate}
For both pNGB DM models (model A and model B), the dominant annihilation proceeds through the Higgs portal, mediated by the CP-even scalars $h_1$ and $h_2$. The relevant channels are
\begin{enumerate}
	\item $\chi^\ast+\chi\to h_{1,2}^\ast \to f+\bar{f}$. Annihilation into SM fermions gives
	\beq
	\langle\sigma_{\bar{f}f} v\rangle&\approx& \sum_{f=t,b,...}\frac{c_f m_f^2}{32\pi v^2 m_{\mathrm{DM}}^2}\left(1-\xi_f^2\right)^{3/2}\nonumber\\
	&&\times\left|\frac{g_{\chi\chi h_1}c_\theta}{4m_{\mathrm{DM}}^2-m_{h_1}^2+im_{h_1}\Gamma_{h_1}}+\frac{g_{\chi\chi h_2}s_\theta}{4m_{\mathrm{DM}}^2-m_{h_2}^2+im_{h_2}\Gamma_{h_2}}\right|^2,
	\eeq
	where $c_f$ is the color factor of the SM fermion $f$. Because the coupling is proportional to the fermion mass squared (inherited from the SM Higgs Yukawa coupling), the dominant contributions arise from the heaviest accessible fermions when $m_{\mathrm{DM}}>m_f$. All relevant scalar couplings were defined in Section~\ref{sec:model}. The total decay width of the heavier state $h_2$ is approximated by
	\beq
	\Gamma_{h_2}\approx\frac{g_{h_1h_1h_2}^2}{32\pi m_{h_2}}\sqrt{1-\frac{4m_{h_1}^2}{m_{h_2}^2}}\Theta(m_{h_2}-2m_{h_1})+\frac{g_{\chi\chi h_2}^2}{16\pi m_{h_2}}\sqrt{1-\frac{4m_{\mathrm{DM}}^2}{m_{h_2}^2}}\Theta(m_{h_2}-2m_{\mathrm{DM}}).
	\eeq
	
	\item {$\chi^\ast+\chi\to V+V$ ($V=W,Z,X$). Annihilation into massive gauge bosons is given by
	\beq
	\langle\sigma_{VV} v\rangle&\approx&\sum_{V=W,Z}\frac{m_{\mathrm{DM}}^2}{8n_V\pi m_V^4}\sqrt{1-\xi_V^2}\left(1-\xi_V^2+\frac{3}{4}\xi_V^4\right)\left|\sum_{i=1}^{2}\frac{g_{\chi\chi h_i}g_{h_iVV}}{4m_{\mathrm{DM}}^2-m_{h_i}^2+im_{h_i}\Gamma_{h_i}}\right|^2\nonumber\\
	&+&\frac{\sqrt{1-\xi_X^2}}{64 \pi m_{\mathrm{DM}}^2}\left[2\left| 2Q_{DM}^2 g_X^2+\sum_{i=1}^2 \frac{g_{\chi \chi h_i} g_{h_i X X}}{4 m_{\mathrm{DM}}^2-m_{h_i}^2+i m_{h_i} \Gamma_{h_i}}\right|^2\right.\nonumber\\
	&&\left.\quad+\left|\frac{2 Q_{DM}^2g_X^2 \xi_X^2}{2-\xi_X^2}+\left(\frac{2-\xi_X^2}{\xi_X^2}\right) \sum_{i=1}^2 \frac{g_{\chi \chi h_i} g_{h_i X X}}{4 m_{\mathrm{DM}}^2-m_{h_i}^2+i m_{h_i} \Gamma_{h_i}}\right|^2\right] \Theta\left(m_{\mathrm{DM}}-m_X\right) ,\nonumber\\
	\eeq
	where $n_V=1(2)$ for the $W(Z)$ boson.
	The coupling between $X_\mu$ and the Higgs fields are given by
	\beq
	g_{h_iXX}=\frac{2m_X^2}{v_\phi}(-s_\theta,c_\theta)\quad (\mathrm{model~A}),\qquad g_{h_iXX}=0\quad (\mathrm{model~B}).
	\eeq
	
	\item $\chi^\ast+\chi\to X+h_i$. Annihilation into $X$ and $h_i$ is also allowed when $2m_{\mathrm{DM}}>m_X+m_{h_i}$. The leading term of annihilation cross section is given by
	\beq
	&&\langle\sigma_{Xh_i}v\rangle \approx	\frac{3Q_{DM}^2g_X^2\beta_{X i} \mathcal{T}_i}{16\pi} x^{-1}\Theta(2m_{\mathrm{DM}}-m_X-m_{h_i})\\
	&&\mathcal{T}_i=A_i^2+	\frac{(m_{\mathrm{DM}} \beta_{X i})^2}{3}\left[\frac{16A_i g_{\chi\chi h_i}}{\Delta_i^2}-\frac{64g_{\chi\chi h_i}^2m_{h_i}^2}{\Delta_i^4}+\frac{g_{h_iXX}^2}{m_X^2(4 m_{\mathrm{DM}}^2-m_X^2)^2}\right],\nonumber
	\eeq
	where 
	\beq
	&&\beta_{X i} \equiv \sqrt{\left[1-\frac{\left(m_X+m_{h_i}\right)^2}{4 m_{\mathrm{DM}}^2}\right]\left[1-\frac{\left(m_X-m_{h_i}\right)^2}{4 m_{\mathrm{DM}}^2}\right]}, \nonumber\\
	&&\Delta_i=4 m_{\mathrm{DM}}^2-m_X^2-m_{h_i}^2,\quad  A_i=\frac{g_{h_iXX}}{4 m_{\mathrm{DM}}^2-m_X^2}-\frac{4g_{\chi\chi h_i}}{4 m_{\mathrm{DM}}^2-m_X^2-m_{h_i}^2}.
	\eeq
	These contributions are also subdominant during freeze-out since they are $p$-wave velocity suppressed.
	}
	
	\item $\chi^\ast+\chi\to h_i+h_j$. When kinematically allowed, annihilation into scalar final states yields
	\beq
	\langle\sigma_{h_ih_j} v\rangle&\approx&\sum_{ij}\frac{1}{32n_{ij}\pi m_{\mathrm{DM}}^2}\sqrt{\left[1-\frac{(m_i+m_j)^2}{4m_{\mathrm{DM}}^2}\right]\left[1-\frac{(m_i-m_j)^2}{4m_{\mathrm{DM}}^2}\right]}\nonumber\\
	&&\times \left|g_{\chi\chi h_i h_j}+\frac{ig_{\chi\chi h_1}g_{h_1h_ih_j}}{4m_{\mathrm{DM}}^2-m_{h_1}^2+im_{h_1}\Gamma_{h_1}}+\frac{ig_{\chi\chi h_2}g_{h_ih_2h_j}}{4m_{\mathrm{DM}}^2-m_{h_2}^2+im_{h_2}\Gamma_{h_2}}\right.\nonumber\\
	&&\left.\quad +\frac{ig_{\chi\chi h_i}g_{\chi\chi h_j}}{m_{h_i}^2-2m_{\mathrm{DM}}^2}+\frac{ig_{\chi\chi h_j}g_{\chi\chi h_i}}{m_{h_j}^2-2m_{\mathrm{DM}}^2}\right|^2,
	\eeq
	where the symmetry factor is $n_{ij}=1+\delta_{ij}$.
\end{enumerate}

{
In addition, since we focus on the case of $m_{\mathrm{DM}}>m_X$ in the secluded dark matter model, we include the Sommerfeld enhancement (SE) induced by repeated $X_\mu$ exchange~\cite{Arkani-Hamed:2008hhe}. The corresponding attractive Yukawa potential in the $\chi^\ast\chi$ initial state is
\beq
V(r)=-\frac{\alpha_\chi}{r}e^{-m_X r},
\eeq
where $\alpha_\chi\equiv (Q_{\mathrm{DM}}g_X)^2/(4\pi)$. We include this effect by multiplying each relevant partial-wave cross section by the corresponding Sommerfeld factor. For the dominant $s$-wave process $\chi^\ast\chi\to XX$,
\beq
(\sigma v)_{XX}^{\mathrm{SE}}= S_0(v_{\rm rel})\,(\sigma v)_{XX}^{\mathrm{tree}},
\eeq
while for the subdominant $p$-wave annihilation into leptons, $(\sigma v)_{\bar{\ell}\ell}^{\mathrm{tree}}=b_\ell v_{\mathrm{rel}}^2$, we use
\beq
(\sigma v)_{\bar{\ell}\ell}^{\mathrm{SE}} = b_\ell v_{\rm rel}^2 S_1(v_{\rm rel}),
\eeq
where $S_1$ denotes the $l=1$ Sommerfeld factor. In the numerical analysis, we evaluate $S_0$ and $S_1$ using the Hulth\'en approximation to the Yukawa potential, following Ref.~\cite{Cassel:2009wt}. More precisely, $S_{0,1}$ below are identified with the $\widetilde S_{0,1}$ factors of Ref.~\cite{Cassel:2009wt}.

The SE-corrected thermal annihilation cross section used in the freeze-out calculation is
\beq
\langle\sigma v\rangle_{\rm sec}^{\mathrm{SE}}(x) = \langle\sigma_{XX}v\rangle_{\rm tree}\,\overline S_0(x) +
\sum_\ell \langle\sigma_{\bar{\ell}\ell}v\rangle_{\rm tree}\,\overline S_1(x),
\label{sigmavSE}
\eeq
where $x=m_{\mathrm{DM}}/T$, $\langle\sigma_{\bar{\ell}\ell}v\rangle_{\rm tree}=6b_\ell/x$, and
\beq
&&\overline{S_0}(x)=\frac{2}{\sqrt{\pi}} \int_0^\infty dt t^{1/2}e^{-t} S_0 \left(2\sqrt{\frac{t}{x}}\right),\\
&&\overline{S_1}(x)=\frac{4}{3\sqrt{\pi}}\int_0^\infty dt\,t^{3/2}e^{-t}S_1 \left(2\sqrt{\frac{t}{x}}\right).
\eeq
Here, the argument of $S_{0,1}$ is the relative velocity, $v_{\rm rel}=2\sqrt{t/x}$. In the notation of Ref.~\cite{Cassel:2009wt}, we take 
$A=\alpha_\chi$, $M=m_{\mathrm{DM}}$, $m_\ast=m_X$, $\delta=\pi^2m_X/6$, and, therefore, $\beta=v_{\rm rel}/2=\sqrt{t/x}$.

We now define the model-dependent thermal annihilation cross section that enters the freeze-out calculation. For the secluded DM model, where $m_{\mathrm{DM}}>m_X$, the Sommerfeld enhancement is included. In this case, we take
\beq
\langle \sigma_{\mathrm{tot}} v\rangle(x)=\langle \sigma v\rangle_{\mathrm{sec}}^{\mathrm{SE}}(x),
\eeq
where $\langle \sigma v\rangle_{\mathrm{sec}}^{\mathrm{SE}}(x)$ is given in Eq.~\eqref{sigmavSE}. 

For the pNGB DM models, after applying the kinematic and perturbativity cuts used in the numerical scan, the displayed viable regions mainly satisfy $m_{\mathrm{DM}}<m_X$ and $\alpha_\chi m_{\mathrm{DM}}/m_X<1$. The Sommerfeld correction from $X_\mu$ exchange is then numerically close to unity. We therefore use the tree-level thermal cross section,
\beq
\langle \sigma_{\mathrm{tot}} v\rangle(x)=\langle\sigma_{\bar{\ell}\ell} v\rangle+\langle\sigma_{\bar{f}f} v\rangle+\langle\sigma_{VV} v\rangle+\sum_i\langle\sigma_{Xh_i} v\rangle+\langle\sigma_{h_i h_j} v\rangle ,
\eeq
where only kinematically open channels are included. 

With this definition, the freeze-out temperature $x_f\equiv m_{\mathrm{DM}}/T_f$ is obtained by iteratively solving
\beq
x_f=\ln\left[\frac{5}{4}\sqrt{\frac{45}{8}}\frac{m_{\mathrm{DM}} M_{\mathrm{Pl}}\langle\sigma_{\mathrm{tot}}v\rangle(x_f)}{2\pi^3\sqrt{g_\ast(x_f)x_f}}\right],
\eeq
where $M_{\mathrm{Pl}}$ is the Planck mass and $g_\ast$ counts the effective relativistic degrees of freedom. The relic abundance is then calculated as
\beq
\Omega_{\mathrm{DM}}h^2\simeq 2.08\times10^9\,{\mathrm{GeV}}^{-1} \left(\frac{T_0}{2.725{\mathrm{K}}}\right)^3
\frac{1}{M_{\mathrm{Pl}}\sqrt{g_\ast(x_f)}J(x_f)},
\label{omegahsq}
\eeq
with
\beq
J(x_f)=\int_{x_f}^{\infty}\frac{dx}{x^2} \langle\sigma_{\mathrm{tot}}v\rangle(x).
\eeq

For channels without sizable Sommerfeld corrections, we can use the standard expansion,
\beq
\langle\sigma_{\mathrm{tot}}v\rangle(x)\simeq a_0+\frac{6a_1}{x}.
\eeq
In this limit,
\beq
J(x_f)=\frac{a_0}{x_f}+\frac{3a_1}{x_f^2},
\eeq
and Eq.~\eqref{omegahsq} reduces to the usual analytic freeze-out expression~\cite{Yu:2011by}. For the secluded model, we instead evaluate $J(x_f)$ using the SE-corrected thermal average in Eq.~\eqref{sigmavSE}.
}

Finally, we consider constraints from indirect detection, primarily driven by $\gamma$-ray observations from Fermi-LAT~\cite{McDaniel:2023bju}. For the secluded DM model, we apply an improved bound given by Ref.~\cite{Su:2025mxv} based on the simulation of $\gamma$-ray spectrum from $\chi^\ast\chi\to XX\to 4\ell$ processes. { Since $m_{\mathrm{DM}}>m_X$ in this case, the Sommerfeld correction is included and evaluated at the characteristic dwarf-galaxy velocity (corresponding to $x\sim 10^{9}$). } 

Note that standard Fermi-LAT limits are typically derived assuming a self-conjugate (Majorana or real scalar) DM candidate. Because our $\chi$ is a complex scalar, the number densities of DM and anti-DM particles are each half of the total DM number density ($n_\chi = n_{\chi^*} = n_{\mathrm{tot}}/2$). Consequently, the annihilation rate, which is proportional to $n_\chi n_{\chi^*}$, acquires a factor of $1/4$ relative to the self-conjugate case (where it is proportional to $n_{\mathrm{tot}}^2/2$). However, because both $\chi^\ast\chi$ and $\chi\chi^\ast$ annihilations contribute to the signal flux, there is a compensating factor of $2$. Combining these effects, we must define an effective annihilation cross section:
\beq
\langle\sigma_{\mathrm{eff}} v\rangle = \frac{1}{2}\langle\sigma_{\mathrm{tot}} v\rangle,
\eeq
which should be used when comparing our model predictions directly against the published Fermi-LAT exclusion bounds. 
	
\subsection{Direct detection and trident production bound}\label{subsect:DD}
For direct detection, we must account for vector-portal scattering induced by the kinetic mixing between the $\Uone_{L_\mu-L_\tau}$ gauge boson $X_\mu$ and the SM hypercharge gauge boson $B_\mu$. Assuming the tree-level kinetic mixing vanishes at the UV scale, an irreducible effective mixing is radiatively generated in the IR via muon and tau loops. This mixing parameter is given by~\cite{Hapitas:2021ilr}:
\beq
&&\mathcal{L}_{mix}=\frac{1}{2}\varepsilon_{IR}X_{\mu\nu}B^{\mu\nu},\\
&&\varepsilon_{IR}(\Lambda<m_\mu)=-\frac{g_X e }{12 \pi^2} \log \left[\frac{m_\tau^2}{m_\mu^2}\right].\label{loopkinmix}
\eeq

The cross section for DM-nucleon ($\chi-N$) scattering is given by~\cite{Koechler:2025ryv}:
\beq
\sigma_N(Q)= \frac{m_N^2m_{\mathrm{DM}}^2}{\pi(m_N+m_{\mathrm{DM}})^2}Q_{\mathrm{DM}}^2g_X^2s_W^2 \varepsilon_{IR}^2(Q) F_{\mathrm{Helm}}^2(Q) \left|\frac{f_N^{(X)}}{m_{X}^2+Q^2}-\frac{f_N^{(Z)}}{\left(m_Z^2-m_{Z^\prime}^2\right)}\right|^2,
\eeq
where the effective nucleon couplings $f_N^{(V)}$ ($V = Z^\prime, Z$) are defined as
\beq
f_N^{(V)}=\frac{1}{A}\left[Z\left(2 g_{u V}+g_{d V}\right)+(A-Z)\left(g_{u V}+2 g_{d V}\right)\right],
\eeq
where $Z'$ is the physical mass eigenstate corresponding to the dark gauge boson $X_\mu$, and the expression of chiral couplings $g_{uV}$ and $g_{dV}$ can be found in Ref.~\cite{Arcadi:2018tly}. In the limit of tiny kinetic mixing ($\varepsilon_{IR} \ll 1$) where $Z'$ is closely aligned with $X_\mu$, the cross section simplifies to
\beq
\sigma_N(Q)\approx  \frac{m_N^2m_{\mathrm{DM}}^2}{\pi(m_N+m_{\mathrm{DM}})^2}\frac{Q_{\mathrm{DM}}^2g_X^2 e^2\varepsilon_{IR}^2}{c_W^2(m_X^2-m_Z^2)^2} \left[(1-2s_W^2)\frac{Z}{A}-\frac{1}{2}\right]^2 F_{\mathrm{Helm}}^2(Q), \label{DDXsect}
\eeq
where $Z=54$ and $A=131$ for xenon-based detectors. At the low momentum transfers typical of direct detection, the Helm form factor is $F_{\mathrm{Helm}}^2(Q) \approx 1$.

{ We also checked direct-detection constraints from electron recoils. In the $U(1)_{L_\mu-L_\tau}$ models considered here, electrons are not charged under the new gauge symmetry at tree level. The coupling of $X_\mu$ to electrons is induced only through the same loop-generated kinetic mixing in Eq.~\eqref{loopkinmix}. The corresponding DM-electron scattering cross section is
\beq
\bar{\sigma}_e\approx\frac{\mu_{\chi e}^2}{\pi}\left[\frac{Q_{\mathrm{DM}}g_X(e\varepsilon_{\mathrm{IR}}/c_W)}{m_X^2}\right]^2.\nonumber
\eeq
For the parameter space considered in this work, $m_\chi\gtrsim100\mathrm{GeV}$, so $\mu_{\chi e}\simeq m_e$. Using Eq.~\eqref{loopkinmix}, this gives
\beq
\bar{\sigma}_e\simeq8.0\times10^{-52}\mathrm{cm}^2\left(\frac{\sqrt{Q_{\mathrm{DM}}}g_X}{0.1}\right)^4\left(\frac{100\mathrm{GeV}}{m_X}\right)^4.
\eeq
Even for the aggressive choice $\sqrt{Q_{\mathrm{DM}}}g_X=0.3$ and $m_X=10~\mathrm{GeV}$, one obtains only $\bar{\sigma}_e\simeq6.5\times10^{-46}~\mathrm{cm}^2$. This is well below the current electron-recoil limits from XENONnT~\cite{XENON:2026qow} and PandaX-4T~\cite{PandaX:2025rrz}, which are at the level of $\bar{\sigma}_e\sim10^{-41}{\mathrm{cm}^2}$ in the sub-GeV to GeV mass range and become weaker when extrapolated to $m_{\mathrm{DM}}\gtrsim100~\mathrm{GeV}$. Therefore, electron-recoil direct detection does not impose additional constraints on the parameter space studied in this work.}

Furthermore, the $\Uone_{L_\mu-L_\tau}$ gauge interactions induce neutrino trident production ($\nu_\mu N \to \nu_\mu \mu^+ \mu^- N$) in accelerator neutrino experiments~\cite{Altmannshofer:2014pba}.
The absence of a significant excess places an experimental constraint on the gauge coupling $g_X$ as a function of $m_X$. As we will show in Section \ref{sec:results}, the direct detection bounds on the DM candidate are typically more restrictive than the trident production constraints in the parameter space of interest.

{ Note that we shall also check the constraint from DM self-interactions. In this work, the transfer cross section induced by $X_\mu$ exchange scales as
\beq
	\frac{\sigma_T}{m_{\mathrm{DM}}} \simeq \frac{m_{\mathrm{DM}}}{4\pi}\left(\frac{Q_{\mathrm{DM}}g_X}{m_X}\right)^4\simeq 1.7\times 10^{-10} \mathrm{ cm^2/g}\left(\frac{Q_{\mathrm{DM}}g_X}{1}\right)^4 \left(\frac{m_{\mathrm{DM}}}{1\mathrm{TeV}}\right)\left(\frac{100\mathrm{GeV}}{m_X}\right)^4 .
\eeq
For the parameter regions we consider in the next section, $m_X$ is at the electroweak or TeV scale, and the resulting $\sigma_T/m_{\mathrm{DM}}$ is far below the commonly used astrophysical bound $\sigma_T/m_{\mathrm{DM}}\lesssim {\cal O}(1)\,{\rm cm^2/g}$ from merging clusters and Bullet-Cluster-like systems~\cite{Markevitch:2003at,Randall:2008ppe,Harvey:2015hha} (see Refs.~\cite{Tulin:2017ara,Adhikari:2022sbh} for recent reviews). Therefore, self-interaction constraints do not further restrict the benchmark parameter space considered in this work.
}

\section{Results}\label{sec:results}
In this section, we apply constraints from direct detection, trident production, indirect detection, and the observed DM relic density to restrict the parameter space of our models. {We also display optimistic maximal-heating sensitivity contours for two benchmark NS configurations. As explained in Section~\ref{subsec:EFT}, these contours correspond to geometric capture saturation supplemented by the assumptions of efficient thermalization and capture-annihilation equilibrium.}

\subsection{Secluded DM model}
In the secluded DM framework, both the direct detection cross section and the NS capture rate depend on the specific parameter combination $\sqrt{Q_{\mathrm{DM}}}g_X$ (see Eqs.~\eqref{effLambda1} and \eqref{DDXsect}). Therefore, it is convenient to fix this combination and present the constraints and sensitivities in the $(\xi_X, m_{\mathrm{DM}})$ plane. 

\begin{figure}
	\centering
	\includegraphics[width=0.45\textwidth]{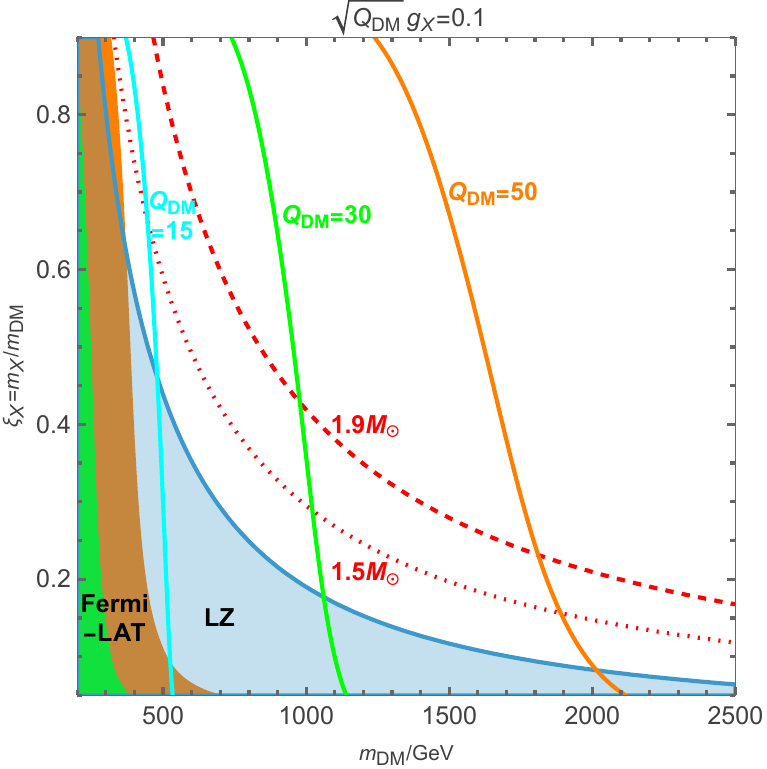}
	\includegraphics[width=0.45\textwidth]{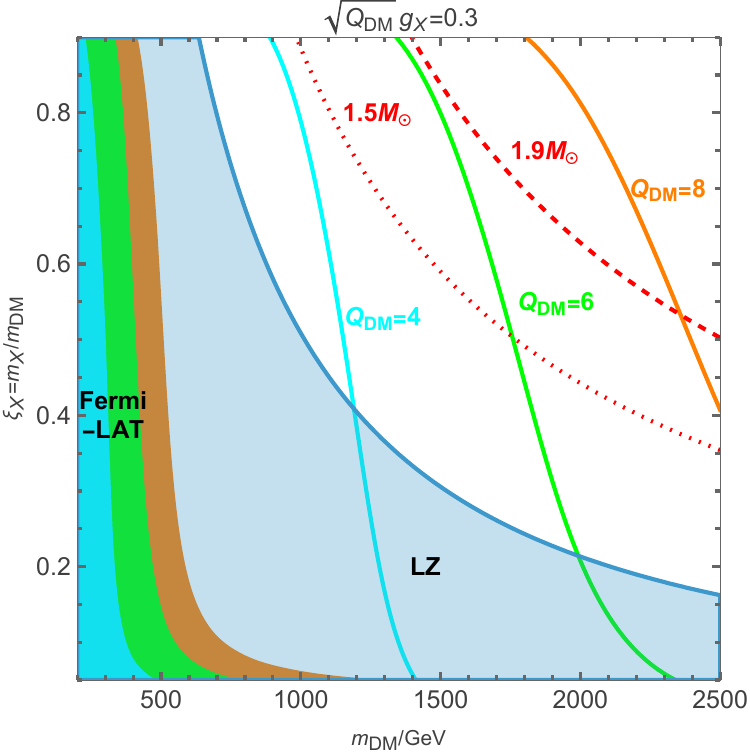}
	\caption{Constraints and prospective sensitivities for the secluded scalar DM model in the $m_{\mathrm{DM}}$--$\xi_X$ plane, with $\sqrt{Q_{\mathrm{DM}}}g_X$ fixed to $0.1$ (left panel) and $0.3$ (right panel). The light blue shaded regions are excluded by the LZ direct detection experiment. The solid cyan, green, and orange curves indicate the parameter space reproducing the correct thermal relic density ($\Omega_{\mathrm{DM}} h^2 \approx 0.12$) for specific charge assignments: $Q_{\mathrm{DM}}=15, 30, 50$ (left) and $Q_{\mathrm{DM}}=4, 6, 8$ (right). The vertical color shaded bands on the left represent the exclusion limits from Fermi-LAT indirect detection corresponding to each $Q_{\mathrm{DM}}$ value. { The red dotted and dashed curves illustrate the optimistic maximal-heating reach of NS observations for BM-1 ($1.5M_\odot$) and BM-2 ($1.9M_\odot$), respectively. For each benchmark NS, the region below the corresponding red curve lies within the projected NS-heating reach and is expected to be probed by near-infrared observations of old NSs under these assumptions.  }}\label{SeclDM}
\end{figure}

In Fig.~\ref{SeclDM}, we show the results for two benchmark values: $\sqrt{Q_{\mathrm{DM}}}g_X=0.1$ (left panel) and $0.3$ (right panel). The light blue shaded regions are excluded by the direct detection bounds from LUX-ZEPLIN~\cite{LZ:2024zvo}. We also evaluated the constraints from neutrino trident production introduced in Section~\ref{subsect:DD}; however, they are significantly weaker than the LZ limits within this mass regime, and thus, fall well outside the plotted parameter space. The red {dotted (BM-1:~$1.5 M_\odot$) and dashed (BM-2:~$1.9 M_\odot$) curves represent the optimistic maximal-heating reach of DM capture by NSs with different masses. Throughout the figures, points below a red NS-heating contour correspond to stronger effective DM-lepton scattering than the saturation threshold for that benchmark star, and hence, lie inside the optimistic maximal-heating reach.} The cyan, green, and orange curves trace the parameter space where the correct relic density ($\Omega_{\mathrm{DM}} h^2 \approx 0.12$) is reproduced for three different charge assignments $Q_{\mathrm{DM}}$ (as labeled near the curves). Additionally, the vertical shaded bands on the left side of the panels indicate regions excluded by the Fermi-LAT indirect detection bounds~\cite{McDaniel:2023bju,Su:2025mxv} for the corresponding charge assignments.

The left panel shows that for a smaller $\sqrt{Q_{\mathrm{DM}}}g_X = 0.1$, a relatively large charge $Q_{\mathrm{DM}} \sim \mathcal{O}(10)$ is required to reproduce the correct relic density. Conversely, when $\sqrt{Q_{\mathrm{DM}}}g_X$ increases to $0.3$ (right panel), smaller charges $Q_{\mathrm{DM}} \sim \mathcal{O}(1)$ become viable.
{In this model, DM annihilation is dominated by gauge interactions. 
For $\xi_X=m_X/m_{\mathrm{DM}}\ll1$, the phase-space factor in $\chi^\ast\chi\to XX$ approaches unity, and the tree-level $s$-wave cross section scales as
\beq
\langle\sigma v\rangle_{XX}^{\mathrm{tree}}\propto \frac{(Q_{\mathrm{DM}}g_X)^4}{m_{\mathrm{DM}}^2}.\nonumber
\eeq
Therefore, without Sommerfeld enhancement, the correct relic density condition would approach a nearly vertical line in the $m_{\mathrm{DM}}$--$\xi_X$ plane at small $\xi_X$. The same scaling also explains the main shape of the Fermi-LAT exclusion bands. For fixed $\sqrt{Q_{\mathrm{DM}}}g_X$ and $Q_{\mathrm{DM}}$, decreasing $m_{\mathrm{DM}}$ increases the present-day annihilation cross section, so Fermi-LAT excludes the low-$m_{\mathrm{DM}}$ region. Since the tree-level cross section depends only weakly on $\xi_X$ when $\xi_X\ll1$, the corresponding Fermi-LAT boundary is also nearly vertical in this region.
	
The Sommerfeld enhancement introduces an additional dependence on the ratio $\alpha_\chi/\xi_X$, where $\alpha_\chi\equiv\frac{(Q_{\mathrm{DM}}g_X)^2}{4\pi}$. For fixed $\sqrt{Q_{\mathrm{DM}}}g_X$, one has
\beq
	Q_{\mathrm{DM}}g_X=\sqrt{Q_{\mathrm{DM}}}\,(\sqrt{Q_{\mathrm{DM}}}g_X),\qquad \alpha_\chi=\frac{Q_{\mathrm{DM}}(\sqrt{Q_{\mathrm{DM}}}g_X)^2}{4\pi}.\nonumber
\eeq
Thus, larger $Q_{\mathrm{DM}}$ gives both a larger annihilation rate and a stronger Sommerfeld correction. As a result, the relic density curves with larger $Q_{\mathrm{DM}}$ are shifted to larger $m_{\mathrm{DM}}$ and show a stronger bending with $\xi_X$. The Fermi-LAT exclusion bands follow the same tendency: When $\xi_X$ becomes smaller, the lighter mediator gives a larger smooth Sommerfeld enhancement, so the excluded region can extend slightly toward larger $m_{\mathrm{DM}}$. By contrast, the lower-mass relic density curves correspond to smaller $Q_{\mathrm{DM}}$, for which the Sommerfeld correction is weaker. These curves, therefore, remain closer to the vertical small-$\xi_X$ asymptote.

The resonant Sommerfeld enhancement approximately occurs at
\beq
\xi_X^{(n)}\simeq \frac{6\alpha_\chi}{\pi^2 n^2},\qquad n=1,2,\ldots .\nonumber
\eeq
For the benchmark couplings used in Fig.~\ref{SeclDM}, the first resonance lies below the displayed range of $\xi_X$. Therefore, the relic density curves and Fermi-LAT exclusion bands shown in Fig.~\ref{SeclDM} are not caused by a narrow Sommerfeld resonance. Their shapes are controlled mainly by the tree-level scaling above, with a smooth nonresonant Sommerfeld correction.
}

Following the improved capture formalism with a relativistic treatment of the scattering kinematics~\cite{Bell:2020lmm}, our inferred NS-heating sensitivity can differ from earlier estimates based on nonrelativistic approximations such as Ref.~\cite{Garani:2019fpa}. In particular, for relativistic and degenerate leptonic targets, this treatment can enhance the capture efficiency, and thus, strengthen the projected reach.
As demonstrated in both panels of Fig.~\ref{SeclDM}, {the optimistic NS heating reach extends beyond the current LZ exclusion region.} This indicates that near-infrared observations of NSs provide a valuable complementary probe for testing leptophilic dark matter models that are otherwise challenging to constrain via terrestrial experiments.

	\subsection{pNGB DM model A}
	As discussed in Section~\ref{subsect:pNGBA}, we assume the $\Uone_{L_\mu-L_\tau}$ gauge symmetry originates from the spontaneous breaking of an $\SUtwo_X$ symmetry in the UV. This theoretically motivates pinning the charges of the doublet components to $Q_{\mathrm{DM}} = -Q_2 = 1/2$, effectively reducing the number of free parameters. The remaining independent parameters are chosen as
	\beq
	v_\phi,\quad m_{h_2},\quad m_X,\quad m_{\mathrm{DM}},\quad t_\theta.
	\eeq
	Note that in minimal scalar-Higgs mixing models, the couplings of the SM-like state $h_1$ to SM fields are universally rescaled by $c_\theta$. Neglecting additional (invisible or exotic) Higgs decays, this implies an approximate rescaling of Higgs signal strengths $\mu\simeq c_\theta^2 \le 1$. Using the ATLAS Run-2 combined global signal strength $\mu=1.06\pm0.06$~\cite{ATLAS-CONF-2021-053}, one finds $c_\theta^2\gtrsim 0.96$ (one-sided 95\% confidence level). Therefore, we adopt the scan choice $t_\theta \lesssim 0.2$.
	
	In terms of these physical parameters, the fundamental couplings $\lambda_{H}$, $\lambda_{\phi}$, $\lambda_{H\phi}$, and $g_X$ can be analytically expressed as
	\beq
	&&\lambda_{H}=\frac{1}{4v^2}\left[m_{h_1}^2+m_{h_2}^2-\frac{m_{h_2}^2-m_{h_1}^2}{\sqrt{1+t_{2\theta}^2}}\right],\quad\lambda_\phi=\frac{1}{4 v_\phi^2}\left[m_{h_1}^2+m_{h_2}^2+ \frac{m_{h_2}^2-m_{h_1}^2}{\sqrt{1+t_{2 \theta}^2}}\right]\nonumber\\
	&&\lambda_{H \phi}=\frac{1}{4 v v_\phi}\left[\frac{(m_{h_2}^2-m_{h_1}^2)t_{2 \theta}}{\sqrt{1+t_{2 \theta}^2}}\right],\quad
	g_X=\frac{2m_X}{v_\phi},
	\eeq
	where $t_{2\theta}=2t_\theta/(1-t_\theta^2)$ and $m_{h_1}\approx 125$~GeV is the mass of the SM-like Higgs boson. 
	
	\begin{figure}
		\centering
		\includegraphics[width=0.45\textwidth]{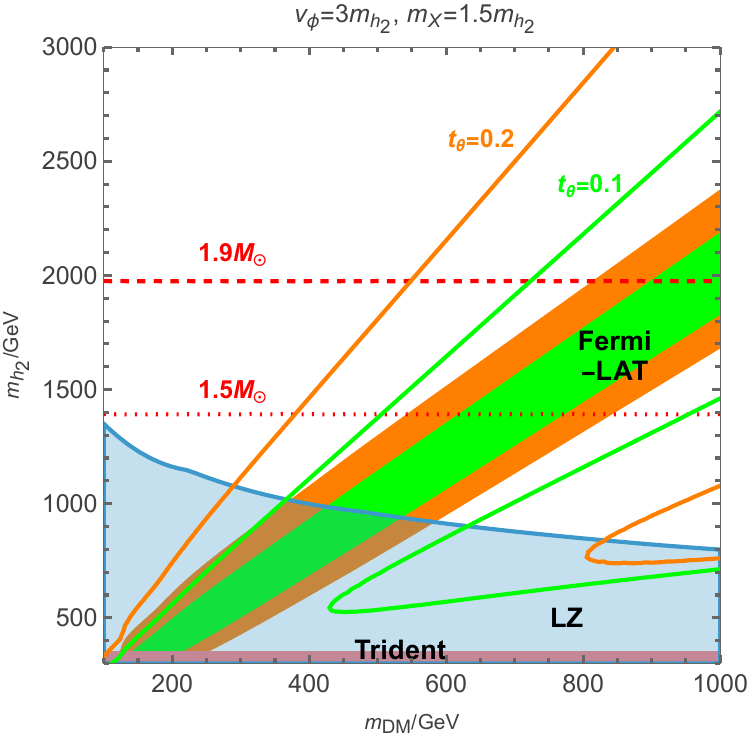}
		\includegraphics[width=0.45\textwidth]{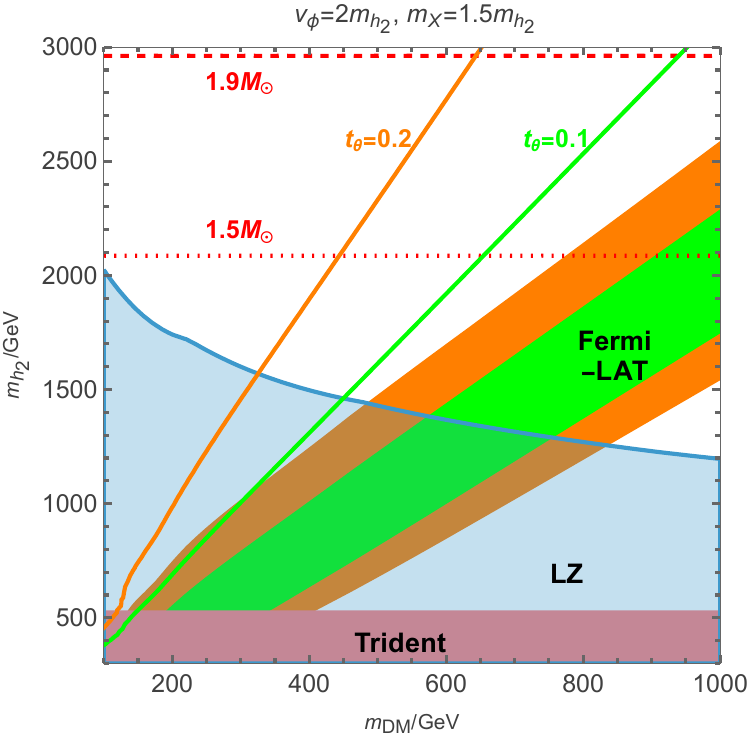}
		\includegraphics[width=0.45\textwidth]{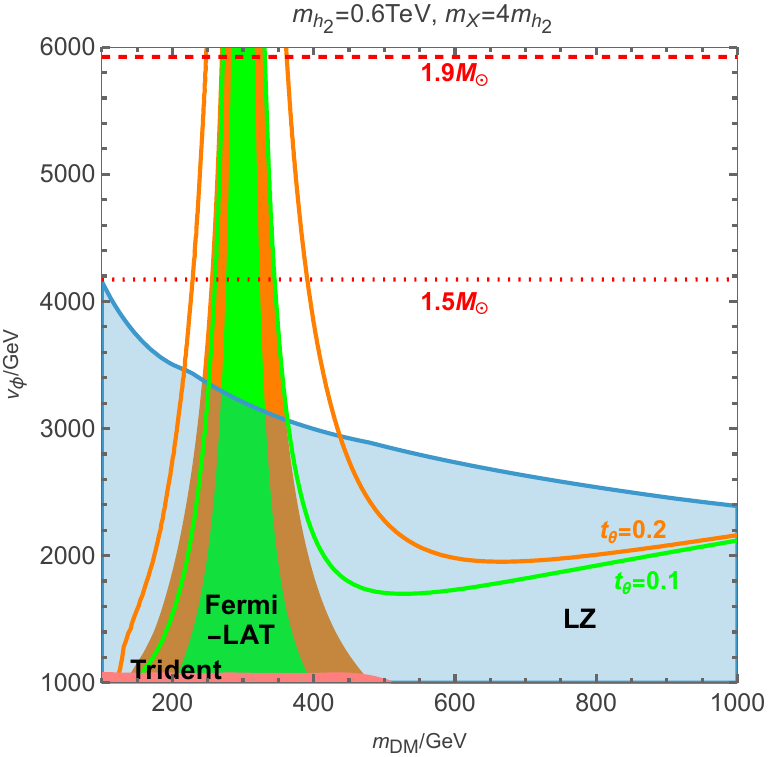}
		\caption{Constraints and prospective sensitivities for pNGB DM model A. The top panels display the results in the $m_{\mathrm{DM}}$--$m_{h_2}$ plane with $m_X=1.5m_{h_2}$, fixing $v_\phi=3m_{h_2}$ (left) and $v_\phi=2m_{h_2}$ (right). The bottom panel projects the constraints onto the $m_{\mathrm{DM}}$--$v_\phi$ plane for fixed masses $m_{h_2}=0.6$~TeV and $m_X=4m_{h_2}$. The light blue and pink shaded regions represent the parameter space excluded by the LZ direct detection and neutrino trident production experiments, respectively. {The solid green and orange curves indicate the regions reproducing the correct thermal relic density ($\Omega_{\mathrm{DM}} h^2 \approx 0.12$) for scalar mixing parameters $t_\theta=0.1$ and $0.2$, respectively. The corresponding color shaded bands denote the regions excluded by Fermi-LAT indirect detection limits for each $t_\theta$. The red dotted and dashed curves illustrate the optimistic maximal-heating reach of NS observations for BM-1 ($1.5M_\odot$) and BM-2 ($1.9M_\odot$), respectively. The parameter space below the corresponding red curve has a sufficiently large DM-lepton interaction to reach this maximal-heating benchmark and is, therefore, expected to be testable by NS-heating observations.}}\label{pNGBMA_fig}
	\end{figure}
	
	In the top row of Fig.~\ref{pNGBMA_fig}, we present the experimental constraints and sensitivities in the $m_{\mathrm{DM}}-m_{h_2}$ plane. Both top panels fix the mass ratio $m_X=1.5m_{h_2}$, with the left and right panels setting $v_\phi=3m_{h_2}$ and $v_\phi=2m_{h_2}$, respectively. The pink shaded regions are excluded by neutrino trident production~\cite{Altmannshofer:2014pba}, while the other colored regions and lines share the same definitions as in the previous subsection. {The orange ($t_\theta=0.2$) and green ($t_\theta=0.1$) curves indicate the parameter space that reproduces the correct DM relic density.} Since the mixing angle $\theta$ roughly scales with the Higgs portal coupling $\lambda_{H\phi}$, a moderately large $t_\theta$ is required to avoid overproducing the relic abundance unless the DM mass lies near the resonance region ($m_{\mathrm{DM}} \approx m_{h_2}/2$). Within the pNGB DM paradigm, such a sizable portal coupling is phenomenologically viable because the tree-level DM-nucleon scattering amplitude is automatically suppressed by the cancellation mechanism.
	
	Nevertheless, the leptophilic gauge interactions inevitably induce a nonzero DM-nucleon scattering cross section via the loop-generated kinetic mixing. From Eqs.~\eqref{loopkinmix} and \eqref{DDXsect}, this vector-portal cross section is approximately proportional to the combination $g_X^4/m_X^4 = 16/v_\phi^4$. This steep quartic dependence on $1/v_\phi$ explains the significant expansion of the LZ exclusion region when shifting from $v_\phi = 3m_{h_2}$ (top left panel) to $v_\phi = 2m_{h_2}$ (top right panel). Furthermore, the neutrino trident production cross section and the NS capture rate exhibit similar parametric dependencies on $g_X$ and $m_X$. Consequently, their corresponding constraints and prospective sensitivities shift simultaneously with the direct detection bounds. In both top panels, we identify viable regions with $m_{\mathrm{DM}}\sim \mathcal{O}(100~\mathrm{GeV})$ that are consistent with all current experimental constraints and fall {within the optimistic maximal-heating reach of future NS observations.}
	
	In the bottom panel of Fig.~\ref{pNGBMA_fig}, we fix $m_{h_2}=0.6$~TeV and $m_X=4m_{h_2}$ as an alternative benchmark, projecting the constraints onto the $m_{\mathrm{DM}}-v_\phi$ plane. This panel confirms that fine tuning the DM mass to the exact resonance region ($m_{\mathrm{DM}} \sim 300$~GeV) is not necessary to reproduce the correct relic density, provided that $t_\theta \gtrsim 0.1$. Finally, this projection demonstrates that NS heating observations have the potential to probe the entire sub-TeV DM parameter space for $v_\phi \lesssim 5.5$~TeV. 
	
	{In all panels of Fig.~\ref{pNGBMA_fig}, the green and orange shaded bands show the Fermi-LAT exclusions for $t_\theta=0.1$ and $0.2$, respectively. These bands lie close to the Higgs portal resonance $m_{h_2}\sim 2m_{DM}$, where the present-day annihilation through the $s$-channel $h_2$ mediator is enhanced. In the displayed parameter space, we mainly focus on $m_X\gtrsim m_{\mathrm{DM}}$ and $\alpha_\chi m_{\mathrm{DM}}/m_X<1$, so the Sommerfeld correction from $X_\mu$ exchange is numerically negligible.}
	
\subsection{pNGB DM model B}
In this model, we assume the $\Uone_{L_\mu-L_\tau}$ gauge field acquires its mass via the Stückelberg mechanism; therefore, $m_X$ is treated as an independent parameter. Since the pNGB DM mass $m_{\mathrm{DM}}$ is generated entirely by gauge interactions at the one-loop level [see Eq.~\eqref{DMmass_loop}], the effective gauge coupling can be expressed as
\beq
Q_{\mathrm{DM}} g_X=\frac{4\pi}{\sqrt{6\ln(r_{UV})}}\frac{m_{\mathrm{DM}}}{m_X},\label{QDMrUVrel}
\eeq
where $r_{UV}\equiv \Lambda_{UV}/m_X$ defines the ratio of the UV matching scale to the gauge boson mass. Consequently, we identify the following set of independent free parameters:
\beq
v_\phi,\quad m_{h_2},\quad m_X,\quad m_{\mathrm{DM}},\quad r_{UV},\quad Q_{\mathrm{DM}},\quad t_\theta.
\eeq
The scalar couplings $\lambda_i$ are determined similarly to model A, yielding
\beq
&&\lambda_{H}=\frac{1}{4v^2}\left[m_{h_1}^2+m_{h_2}^2-\frac{m_{h_2}^2-m_{h_1}^2}{\sqrt{1+t_{2\theta}^2}}\right],\quad \lambda_{33}=\frac{1}{4 v_\phi^2}\left[m_{h_1}^2+m_{h_2}^2+ \frac{m_{h_2}^2-m_{h_1}^2}{\sqrt{1+t_{2 \theta}^2}}\right]\nonumber\\
&&\lambda_{H 3}=\frac{1}{4 v v_\phi}\left[\frac{(m_{h_2}^2-m_{h_1}^2)t_{2 \theta}}{\sqrt{1+t_{2 \theta}^2}}\right],\quad\lambda_{3\chi}=\frac{\lambda_{33}}{1-\frac{m_{\mathrm{DM}}^2}{m_X^2}},\quad \lambda_{H\chi}=\frac{\lambda_{H3}}{1-\frac{m_{\mathrm{DM}}^2}{m_X^2}}.\label{ren_lambdas}
\eeq

{ In the following numerical analysis, we impose the conservative perturbativity condition	
\beq
Q_{\mathrm{DM}}g_X< \sqrt{\pi}\Leftrightarrow\alpha_\chi<\frac{1}{4}.
\eeq
For $\alpha_\chi>1/4$, the leading-log treatment is regarded as unreliable, and higher order corrections need to be included. For the benchmark choices $r_{UV}=10$ and $10^3$, this condition implies
\beq
\frac{m_{\mathrm{DM}}}{m_X}<\frac{\sqrt{6\ln r_{UV}}}{4\sqrt{\pi}}\simeq 0.52~(0.91),
\eeq
respectively. Therefore, the displayed perturbative benchmark region automatically satisfies $m_{\mathrm{DM}}<m_X$ and does not hit the pole in Eq.~\eqref{ren_lambdas}. }

\begin{figure}
	\centering
	\includegraphics[width=0.45\textwidth]{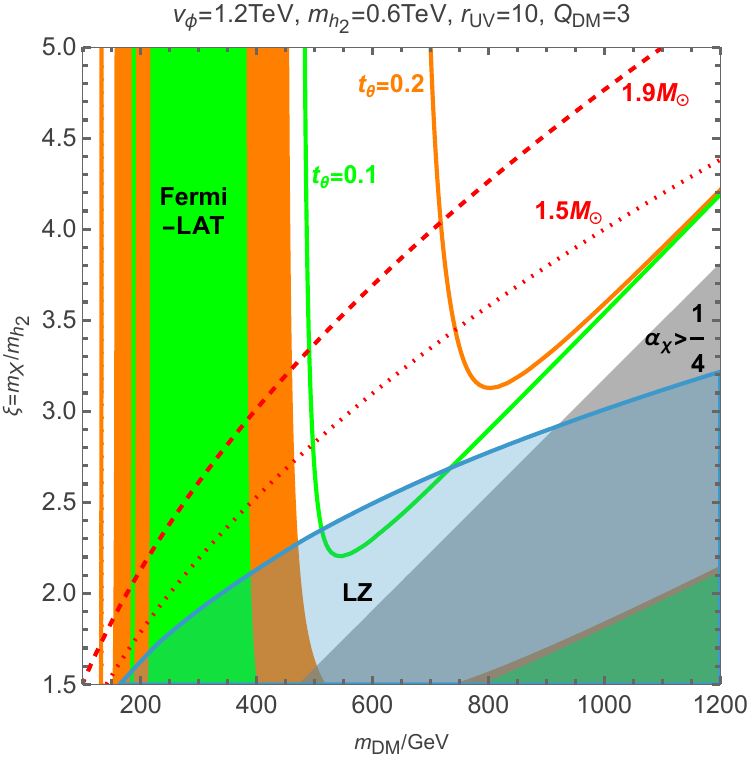}
	\includegraphics[width=0.45\textwidth]{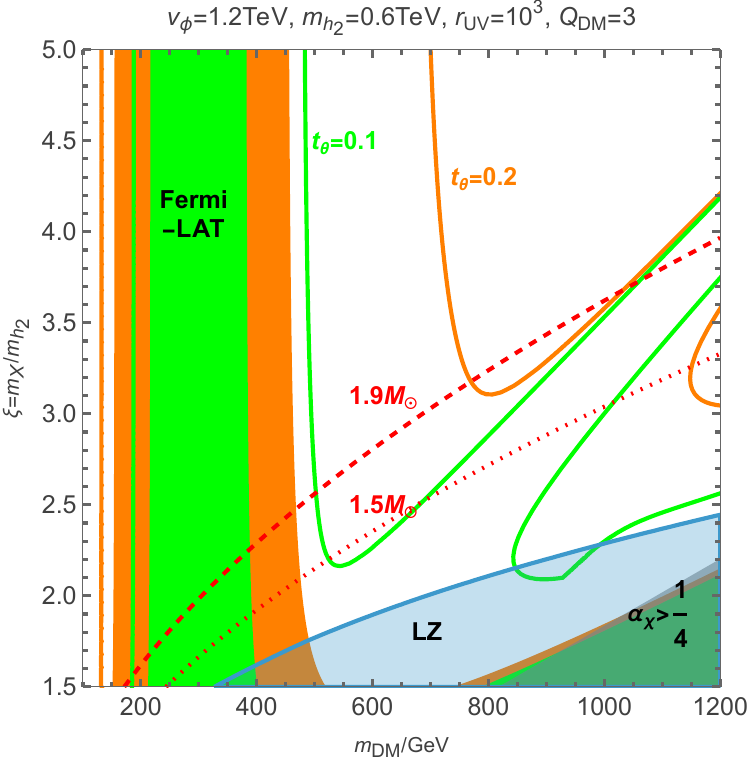}
	\includegraphics[width=0.45\textwidth]{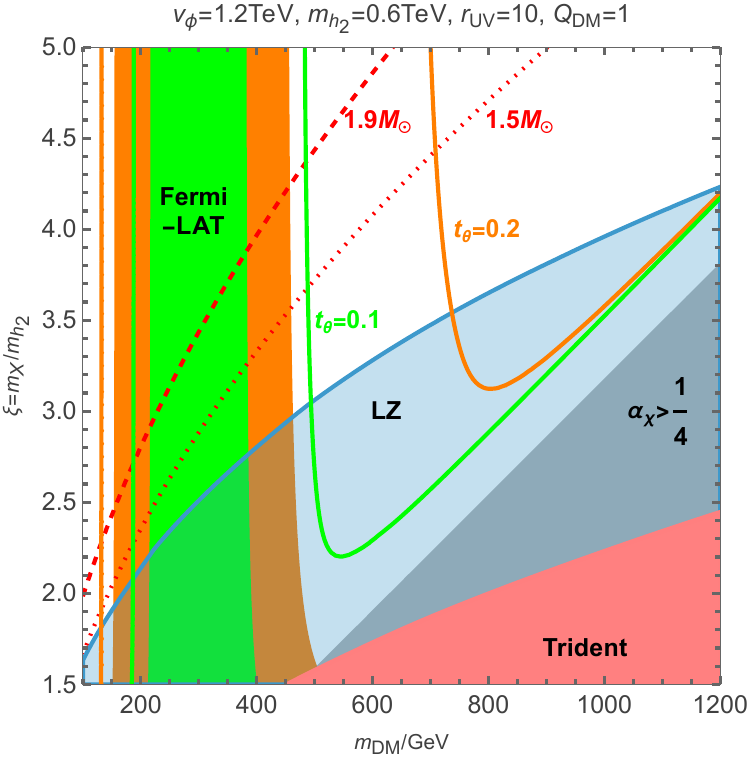}
	\includegraphics[width=0.45\textwidth]{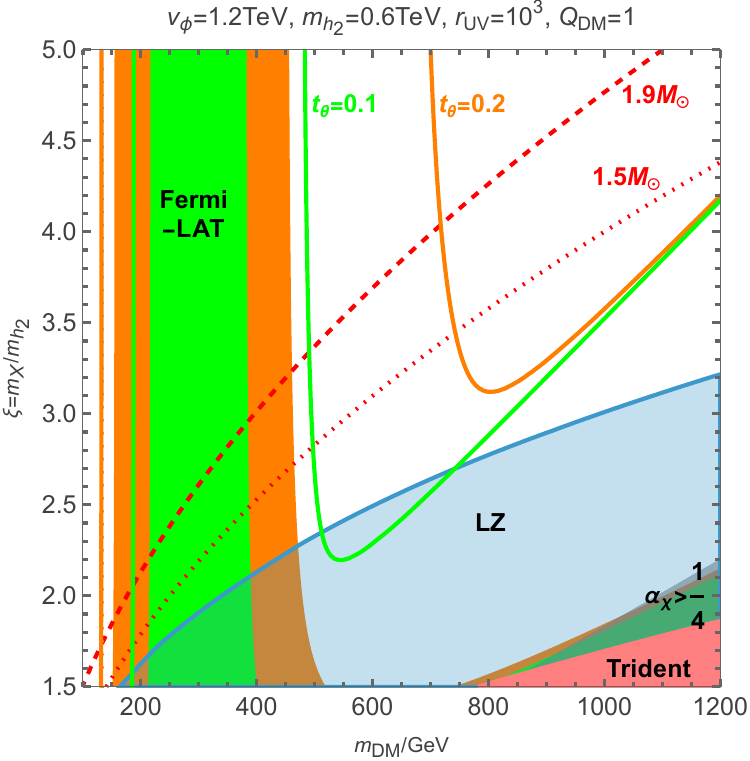}
	\caption{Constraints and prospective sensitivities for pNGB DM model B in the $m_{\mathrm{DM}}$--$\xi$ plane, where the mass ratio is defined as $\xi \equiv m_X/m_{h_2}$. The fixed parameters are $v_\phi=1.2$~TeV and $m_{h_2}=0.6$~TeV across all panels. The four subplots correspond to different combinations of the DM charge $Q_{\mathrm{DM}}$ (top row: $Q_{\mathrm{DM}}=3$; bottom row: $Q_{\mathrm{DM}}=1$) and the UV matching scale ratio $r_{UV}$ (left column: $r_{UV}=10$; right column: $r_{UV}=10^3$). The light blue and pink shaded regions represent the parameter space excluded by the LZ direct detection and neutrino trident production experiments, respectively. {The solid green and orange curves trace the regions reproducing the correct thermal relic density ($\Omega_{\mathrm{DM}} h^2 \approx 0.12$) for scalar mixing parameters $t_\theta=0.1$ and $0.2$, respectively. The corresponding color shaded bands denote the regions excluded by Fermi-LAT indirect detection limits for each $t_\theta$. The red dotted and dashed curves illustrate the optimistic maximal heating reach of NS observations for BM-1 ($1.5M_\odot$) and BM-2 ($1.9M_\odot$), respectively. In each panel, the region below a given red curve is within the corresponding projected NS-heating sensitivity. The gray shaded region labeled $\alpha_\chi>1/4$ is regarded as outside the perturbative benchmark domain.}}\label{figmodelB1}
\end{figure}

In Fig.~\ref{figmodelB1}, we present the experimental constraints and prospective sensitivities in the $m_{\mathrm{DM}}-\xi$ plane, where $\xi \equiv m_X/m_{h_2}$, for four benchmark sets of fixed parameters (labeled above each panel). The light blue and pink shaded regions represent the parameter space excluded by the LZ direct detection and neutrino trident production experiments, respectively. The curves corresponding to the observed relic density for different values of $t_\theta$ clearly converge around two distinct resonance regions: $m_{\mathrm{DM}} \approx m_{h_2}/2 = 300$~GeV and $m_{\mathrm{DM}} \approx m_{X}/2 = \xi \cdot 300$~GeV. The former is driven by $s$-channel DM annihilation into SM particles mediated by $h_2$. The latter corresponds to $s$-channel annihilation mediated by $X_\mu$ into $\mu$ and $\tau$ leptons. Because the vector-mediated annihilation of a scalar DM candidate is a $p$-wave process, its cross section is heavily velocity suppressed in the present-day Universe compared to the epoch of DM freeze-out. As a result, the resonance region around $m_{\mathrm{DM}} \approx m_{X}/2$ is effectively unconstrained by current indirect detection bounds. { The gray shaded regions denoted as $\alpha_\chi>1/4$ are excluded by the conservative perturbativity bound. }

\begin{figure}
	\centering
	\includegraphics[width=0.45\textwidth]{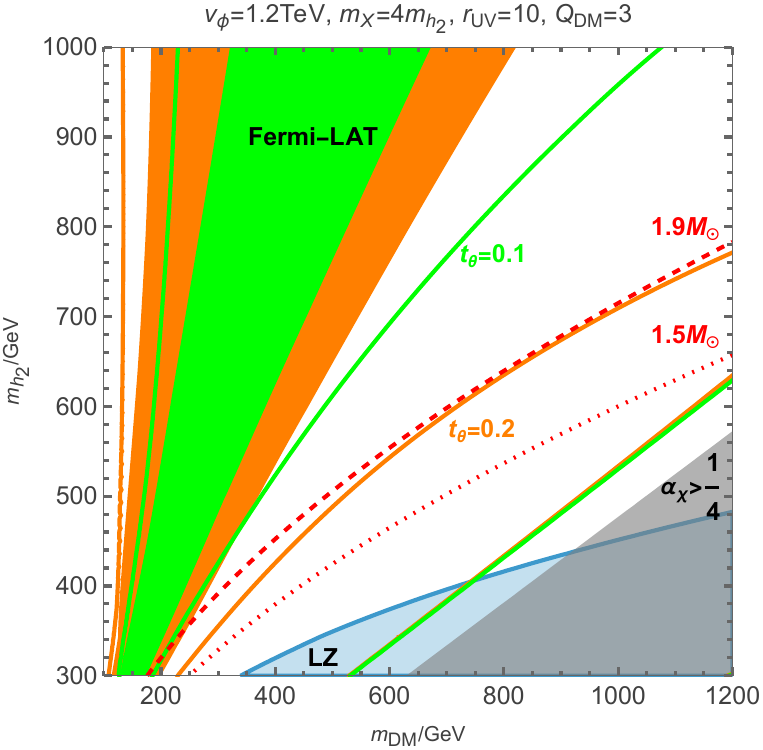}
	\includegraphics[width=0.45\textwidth]{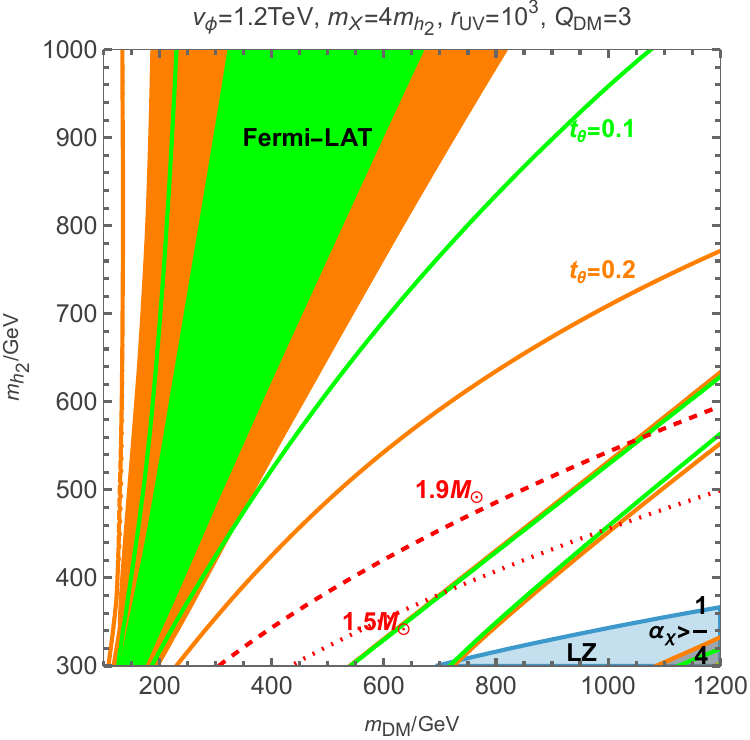}
	\includegraphics[width=0.45\textwidth]{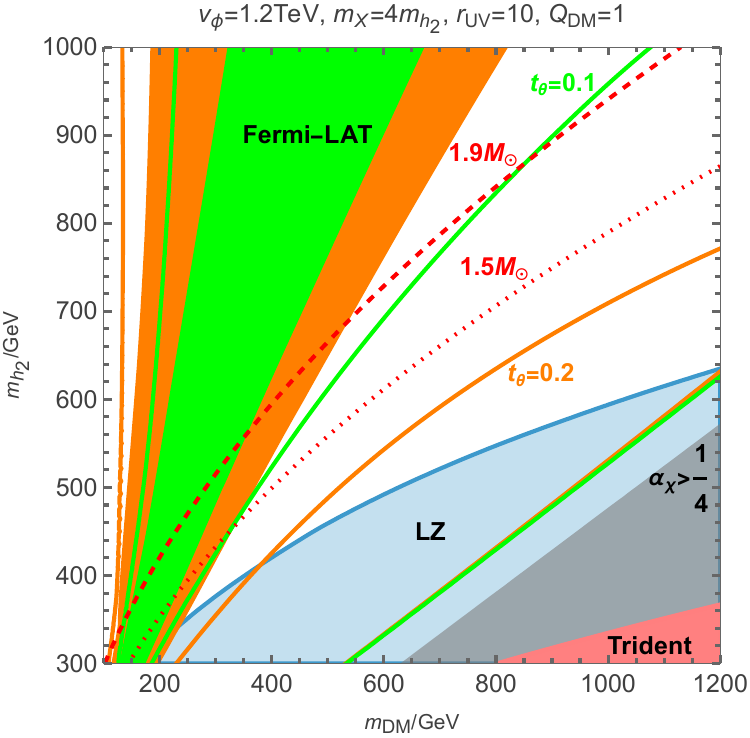}
	\includegraphics[width=0.45\textwidth]{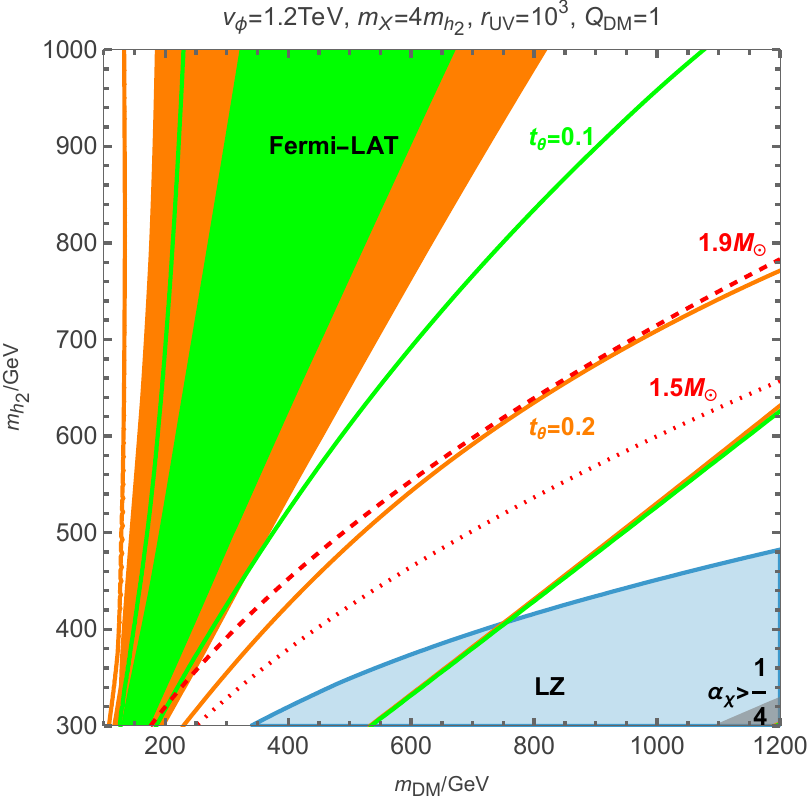}
	\caption{Constraints and prospective sensitivities for pNGB DM model B projected onto the $m_{\mathrm{DM}}$--$m_{h_2}$ plane. The gauge boson mass is fixed to $m_X=4m_{h_2}$, and the VEV is set to $v_\phi=1.2$~TeV across all panels. Similar to Fig.~\ref{figmodelB1}, the panels represent four benchmark combinations of $Q_{\mathrm{DM}}$ (three for the top row, one for the bottom row) and $r_{UV}$ (10 for the left column, $10^3$ for the right column). The light blue and pink shaded regions correspond to exclusions from LZ and neutrino trident production, respectively. {The green and orange curves indicate the correct relic density for $t_\theta=0.1$ and $0.2$, accompanied by their corresponding color Fermi-LAT exclusion bands. The red dotted and dashed curves illustrate the optimistic maximal-heating reach of NS observations for BM-1 ($1.5M_\odot$) and BM-2 ($1.9M_\odot$), respectively. In each panel, the region below a given red curve is within the corresponding projected NS-heating sensitivity. The gray shaded region labeled $\alpha_\chi>1/4$ is regarded as outside the perturbative benchmark domain.}}\label{figmodelB2}
\end{figure}

In Fig.~\ref{figmodelB2}, we show the corresponding constraints and sensitivities projected onto the $m_{\mathrm{DM}}-m_{h_2}$ plane for another four benchmark parameter sets. Consistent with the previous figure, the indirect detection bounds appear near the Higgs portal resonance region ($m_{h_2} \sim 2m_{\mathrm{DM}}$) but are notably absent near the vector-mediated resonance region ($m_X \sim 2m_{\mathrm{DM}}$). { The gray shaded regions denoted as $\alpha_\chi>1/4$ are excluded by the conservative perturbativity bound. }

To systematically understand the variations across the four panels in both Figs.~\ref{figmodelB1} and \ref{figmodelB2}, it is instructive to analyze the dependence of the observables on the parameters $r_{UV}$ and $Q_{\mathrm{DM}}$. From Eq.~\eqref{QDMrUVrel}, the effective parameter combination governing the vector-portal scattering amplitudes can be expressed as $\sqrt{Q_{\mathrm{DM}}}g_X = (Q_{\mathrm{DM}} g_X)/\sqrt{Q_{\mathrm{DM}}}$. For a given kinematic ratio $m_{\mathrm{DM}}/m_X$, this combination scales as $[\ln(r_{UV})]^{-1/2} Q_{\mathrm{DM}}^{-1/2}$. { Consequently, increasing $r_{UV}$ from the left column to the right column decreases $\sqrt{Q_{\mathrm{DM}}}g_X$, which weakens the LZ exclusion and the optimistic NS-heating reach. Lowering the charge from $Q_{\mathrm{DM}}=3$ to $Q_{\mathrm{DM}}=1$ has the opposite effect and makes these two probes more sensitive. This also explains the similarity between the top left and bottom right panels in Figs.~\ref{figmodelB1} and \ref{figmodelB2}, since $Q_{\mathrm{DM}}\ln r_{UV}$ is the same for these two benchmark choices. The trident bound depends directly on $g_X=(Q_{\mathrm{DM}}g_X)/Q_{\mathrm{DM}}$, and is, therefore, stronger in the bottom right panels than in the top left panels. By contrast, the perturbativity bound depends on $Q_{\rm DM}g_X$, so it is weaker in the bottom right panels.}

In contrast, the curves for the thermal relic density and the Fermi-LAT indirect detection bounds exhibit minimal variation with respect to $r_{UV}$ and $Q_{\mathrm{DM}}$, except in the immediate vicinity of the vector-portal resonance ($m_X \approx 2m_{\mathrm{DM}}$). This stability arises because, away from the gauge resonance, the DM annihilation processes are dominated by the Higgs portal interactions, rendering them largely insensitive to variations in the gauge sector parameters. Ultimately, since the Higgs portal DM-nucleon scattering is automatically suppressed by the pNGB mechanism, significant portions of the parameter space with $m_{\mathrm{DM}} \sim \mathcal{O}(100~\mathrm{GeV})$ and $t_\theta \gtrsim 0.1$ remain viable under all current experimental limits. {The benchmark results show that sizable regions of parameter space survive all current constraints. In these regions, the optimistic projected sensitivity from NS heating can provide an independent complementary probe. 

Before closing this subsection, we comment on the capture-annihilation equilibrium assumption in the vector-resonance region of model B. Around $m_X\simeq 2m_{\mathrm{DM}}$, the relic density contours can be strongly affected by the $X_\mu$-mediated process $\chi^\ast\chi\to X^\ast\to \ell^+\ell^-$, which is $p$ wave for scalar dark matter. After capture and thermalization inside an old NS, this vector-mediated channel is, therefore, velocity suppressed. However, this does not imply that annihilation heating is absent in this region. Along the displayed $t_\theta=0.1$ and $0.2$ relic density contours, the Higgs portal $s$-wave channels into electroweak gauge bosons and scalar pairs remain open and typically give
\beq
\langle\sigma v\rangle_s \sim 10^{-27}-10^{-25}\mathrm{cm}^3/\mathrm{s}
\eeq
near $m_X\simeq 2m_{\rm DM}$. Reference~\cite{Bell:2023ysh} has shown that, for broad classes of $s$-wave-annihilating WIMP-like DM, efficient capture is accompanied by capture-annihilation equilibrium well within the age of an old NS. We therefore expect these benchmark regions to satisfy the capture-annihilation equilibrium assumption relevant for maximal NS heating. The red NS-heating curves should, nevertheless, still be interpreted as optimistic reach contours, since they also assume geometric capture saturation and efficient thermalization.
}

	\section{Summary and discussion}\label{sec:summary}
	In this work, we have comprehensively investigated three representative leptophilic scalar dark matter models designed to evade the stringent constraints from current direct detection experiments. Our primary objective is to evaluate the prospective sensitivity of near-infrared (near-IR) telescopes to dark matter by observing the thermal emission from old neutron stars. The baseline framework extends the Standard Model with a $\Uone_{L_\mu-L_\tau}$ gauge symmetry. Within this setup, WIMP dark matter charged under this symmetry can be efficiently captured by neutron stars via scattering off the degenerate muon population, mediated by the $\Uone_{L_\mu-L_\tau}$ gauge boson $X_\mu$. {If the captured DM population thermalizes efficiently and reaches capture-annihilation equilibrium, the kinetic energy deposited during capture together with the annihilation contribution can produce a maximal-heating signal with a surface temperature of order $2000~\mathrm{K}$.}
	
	First, to establish a model-independent baseline, we analyzed the neutron star capture rate using a dimension-6 effective operator (Eq.~\eqref{DMlepscattEff}) describing the DM-lepton interaction. Assuming two benchmark neutron star masses, $M_\star=1.5 M_\odot$ (BM-1) and $1.9 M_\odot$ (BM-2), we computed the corresponding DM capture rates. {By requiring the capture process to reach the geometric limit, we determined the capture-saturation threshold for the effective cutoff scale to be $\Lambda_\ast\approx 2.95$~TeV and $4.19$~TeV for BM-1 and BM-2, respectively. These thresholds enter our optimistic maximal-heating reach.}
	
	Second, we constructed three concrete models that realize this effective interaction. The first is a secluded scalar dark matter model, where the Higgs portal coupling is assumed to be negligible, thereby naturally suppressing tree-level DM-nucleon scattering. The other two models are formulated within the pNGB dark matter paradigm, which automatically suppresses the zero-momentum-transfer scattering amplitude via an approximate shift symmetry. These two pNGB models are distinguished by their mass generation mechanisms: in model A, based on an approximate global $\SO(4)$ symmetry, the pNGB mass is generated by a tree-level soft-breaking mass term; in model B, based on an $\SO(3)$ symmetry, the pNGB mass is radiatively generated at the one-loop level via the $\Uone_{L_\mu-L_\tau}$ gauge interactions.
	
	Finally, we derived the relevant DM annihilation cross sections and the loop-induced DM-nucleon scattering cross sections to evaluate the phenomenology of these models. Our results indicate that all three models successfully accommodate a DM candidate with a mass $m_{\mathrm{DM}}\sim\mathcal{O}(100)~\GeV$. Furthermore, the { optimistic} projected sensitivity of neutron star heating observations can probe regions of the parameter space that remain inaccessible to current terrestrial direct detection experiments. In particular, for the pNGB DM models, the cancellation mechanism disentangles the interactions responsible for direct detection from those dictating the relic density. Consequently, the correct relic abundance can be achieved without strictly requiring the DM mass to reside near a mediator resonance.
	
	We also note that these gauged $\Uone_{L_\mu-L_\tau}$ models can be independently tested at future muon collider experiments. The dominant channels for $X_\mu$ searches are the $2 \to 2$ process $\mu^++\mu^-\to \gamma,Z,X\to \ell^++\ell^-$ and the radiative process $\mu^++\mu^-\to \gamma,Z,X\to \ell^++\ell^-+\gamma$. For a muon collider operating at a center-of-mass energy $\sqrt{s}=3$~TeV with an integrated luminosity of $1~\textrm{ab}^{-1}$, the $2\sigma$ sensitivity has been estimated in Ref.~\cite{Huang:2021nkl}. Their analysis shows that the sensitivity to the gauge coupling $g_X$ can reach $\mathcal{O}(10^{-3})$. {Since the projected muon-collider sensitivity can reach values of $g_X$, about an order of magnitude smaller than the couplings relevant for our relic density and NS-heating benchmark regions, future muon colliders can provide an independent test of this framework.}
	
	\begin{acknowledgments}
		During the preparation of this manuscript, we used ChatGPT and Gemini for language polishing and consistency checking. The authors retain full responsibility for the content, calculations, figures, and conclusions presented in this work. 
		C. C. is supported by the Guangzhou Science and Technology Planning Project under Grant No.~2023A04J0008, Guangdong Provincial Key Laboratory of Gamma-Gamma Collider and Its Comprehensive Applications, and Guangdong Provincial Key Laboratory of Advanced Particle Detection Technology (2024B1212010005). H.-H. Z. is supported by the National Natural Science Foundation of China (NSFC) under Grants No.~12275367, the Fundamental Research Funds for the Central Universities, and the Sun Yat-Sen University Science Foundation.
		
	\end{acknowledgments}

	\section*{Data Availability}
	There are no publicly available research data or software supporting this manuscript. Requests for further information or data should be sent to the authors.
	
	\appendix
	{
	\section{Covariant derivation of the interaction rate}
	\label{app:int_rate}
	
	In this Appendix, we present a self-contained derivation of the relativistic interaction rate $\Omega^-(r)$ used in the neutron-star capture calculation. The purpose is to make explicit the normalization of the flux and phase-space factors in the local rest frame of the neutron-star medium. This is especially	useful for leptonic targets, since electrons and muons inside a neutron star can be relativistic and highly degenerate. Our derivation follows the spirit of	the relativistic capture formalism developed in Refs.~\cite{Bell:2020obw,Bell:2020jou}, while keeping the M{\o}ller flux factor and the final-state Pauli-blocking factor explicit.
	
	We work in the local inertial frame comoving with the neutron-star medium at a	fixed radius $r$. The incoming DM particle and the initial target lepton have four-momenta
	\beq
	k^\mu=(E_\chi,\mathbf{k}),\qquad p^\mu=(E_\ell,\mathbf{p}),
	\eeq
	with 
	\beq
	E_\chi^2=p_\chi^2+m_\chi^2,\qquad E_\ell^2=p_\ell^2+m_\ell^2 .
	\eeq
	In this Appendix, we use $m_\chi=m_{\mathrm{DM}}$ and $m_\ell$ to denote the mass of dark matter and lepton.
	
	For halo DM falling into the neutron-star gravitational potential, neglecting the small velocity at infinity, the local DM energy and momentum are
	\beq
	E_\chi(r)=\frac{m_\chi}{\sqrt{B(r)}},\qquad	p_\chi(r)=m_\chi\sqrt{\frac{1-B(r)}{B(r)}} .
	\eeq
	The final-state momenta are denoted by
	\beq
	k'^\mu=(E'_\chi,\mathbf{k}'),\qquad	p'^\mu=(E'_\ell,\mathbf{p}').
	\eeq
	For convenience, we define
	\beq
	\beta(s)\equiv s-m_\chi^2-m_\ell^2,\qquad \gamma(s)\equiv \lambda^{1/2}(s,m_\chi^2,m_\ell^2) = \sqrt{\beta^2(s)-4m_\chi^2m_\ell^2},
	\eeq
	where $	s=(k+p)^2,~t=(k-k')^2=(p'-p)^2 $ are the Mandelstam variables. 	The M{\o}ller flux factor entering the local interaction rate is
	\beq
	v_{\rm M{\o}l}=	\frac{\sqrt{(k\cdot p)^2-m_\chi^2m_\ell^2}}{E_\chi E_\ell}=	\frac{\gamma(s)}{2E_\chi E_\ell}.
	\eeq
	It should be distinguished from the invariant relative speed $v_{\rm rel}=\gamma(s)/\beta(s)$. 
	
	The local interaction rate per incoming DM particle can be written as
	\beq
	\Omega^-(r)	&=&	g_\ell \zeta_\ell(r) \int\frac{d^3p}{(2\pi)^3}	f_{\rm FD}(E_\ell,r) \int d\sigma v_{\rm M{\o}l} \left[1-f_{\rm FD}(E'_\ell,r)\right] \Theta(E'_\ell-E_\ell) ,\label{omega_master}
	\eeq
	where $g_\ell=2$ for the lepton target. The factor $\zeta_\ell(r)\equiv n_\ell(r)/n_\ell^{\rm free}(r)$ accounts for the use of
	the lepton density from the neutron-star equation of state, and for a free degenerate lepton gas, $\zeta_\ell=1$. Note that $\left[1-f_{\rm FD}(E'_\ell,r)\right]$ encodes the Pauli-blocking effect, while $\Theta(E'_\ell-E_\ell)=\Theta(E_\chi-E'_\chi)$ encodes the capture condition. For an old nonrotating neutron star, we can take a zero temperature approximation, which leads to
	\beq
	f_{\rm FD}(E_\ell,r)\to \Theta\left(\mu_\ell(r)-E_\ell\right),\qquad \left[1-f_{\rm FD}(E'_\ell,r)\right] \to \Theta\left(E'_\ell-\mu_\ell(r)\right).
	\eeq
	
	For a $2\to2$ elastic scattering process with a spin-averaged matrix element $\overline{|\mathcal M|^2}$, the azimuth-integrated differential cross section is
	\beq
	\frac{d\sigma}{dt} = \frac{\overline{|\mathcal M|^2}}{16\pi\,\lambda(s,m_\chi^2,m_\ell^2)} = \frac{\overline{|\mathcal M|^2}}{16\pi\,\gamma^2(s)} .	\label{eq:dsigma_dt_app}
	\eeq
	However, the final-state Pauli-blocking factor depends on the final lepton energy in the local rest frame of the neutron-star medium. At fixed $(E_\ell,s,t)$ this energy generally depends on the final-state azimuthal angle $\varphi_\ast$ in the $\chi\ell$ center-of-mass (CM) frame. It is, therefore, useful to keep the two-body final-state phase space differential in $\varphi_\ast$:
	\beq
	\frac{d\sigma}{dt d\varphi_\ast} = \frac{\overline{|\mathcal M|^2}}{32\pi^2 \gamma^2(s)}. \label{eq:dsigma_dtdphi_app}
	\eeq	
	Here, $\varphi_\ast$ is not a local-frame angle, but only the azimuthal coordinate used to label the Lorentz-invariant two-body final-state phase space	in the $\chi\ell$ CM frame. The factor $d\sigma/(dt d\varphi_\ast) dt d\varphi_\ast$ is, therefore, the correct	differential cross section for this phase-space element. For each $\varphi_\ast$, we boost the final lepton momentum back to the local rest frame of the neutron-star medium and evaluate the Pauli-blocking factor using the corresponding local energy $E'_\ell(\varphi_\ast)$. Thus, no separate transformation of the measure $d\varphi_\ast$ is required.
	
	We now rewrite Eq.~\eqref{omega_master} using the variables	$(E_\ell,s,t,\varphi_\ast)$. Choosing the local $z$ axis along the incoming DM	momentum $\mathbf{k}$, one has
	\beq 
	s = m_\chi^2+m_\ell^2+2(E_\chi E_\ell-p_\chi p_\ell\cos\alpha),
	\eeq
	where $\alpha$ is the angle between $\mathbf{k}$ and $\mathbf{p}$. Therefore, for fixed $E_\ell$, the range of $s$ is $s_-\leq s\leq s_+$, where
	\beq
	s_\pm(E_\ell)	&=&	m_\chi^2+m_\ell^2 +2(E_\chi E_\ell\pm p_\chi p_\ell). \label{s_limits}
	\eeq
	At fixed $s$, the elastic momentum-transfer range is $-\gamma^2(s)/s\le t\le0$. 
	
	At fixed $E_\ell$, one has
	\beq
	ds=-2p_\chi p_\ell d\cos\alpha .
	\eeq
	Thus, after reversing the integration limits from $\cos\alpha\in[-1,1]$ to $s\in[s_-(E_\ell),s_+(E_\ell)]$, the target momentum	measure becomes
	\beq
	d^3p\longrightarrow	\frac{E_\ell}{2p_\chi} dE_\ell ds d\phi,
	\eeq
	and the integral of $\phi$ gives a factor $2\pi$. In the zero-temperature limit, the initial-state factor restricts $E_\ell\le \mu_\ell(r)$. Therefore, for DM falling from rest at infinity, the final-state condition $E'_\ell>\mu_\ell(r)$ automatically implies the capture condition $E'_\ell>E_\ell$. Combining the above, the local interaction rate becomes
	\beq
	\Omega^-(r)=\frac{g_\ell\,\zeta_\ell(r)}{256\pi^3\,p_\chi(r)E_\chi(r)}	\int_{m_\ell}^{\mu_\ell(r)}dE_\ell \int_{s_-(E_\ell)}^{s_+(E_\ell)}ds	\int_{-\gamma^2(s)/s}^{0}dt	\frac{\overline{|\mathcal M|^2}(s,t)}{\gamma(s)}\overline{\mathcal P}_\ell(E_\ell,s,t;r)	\label{eq:omega_final_app}
	\eeq
	where 
	\beq
	\overline{\mathcal P}_\ell(E_\ell,s,t;r) = \frac{1}{2\pi}\int_0^{2\pi}d\varphi_\ast  \Theta\left(E'_\ell(\varphi_\ast)-\mu_\ell(r)\right). \label{eq:Pbar_def_app}
	\eeq
	Its meaning is the fraction of $\varphi_\ast$ for which $E'_\ell>\mu_\ell(r)$.
	
	In the CM frame, we take the polar axis along the initial lepton momentum,
	\beq
	E_\ell^\ast = \frac{s+m_\ell^2-m_\chi^2}{2\sqrt{s}}, \qquad	p_\ast = \frac{\gamma(s)}{2\sqrt{s}},
	\eeq
	and the Lorentz factor and velocity of the CM frame with respect to the local medium frame are
	\beq
	\Gamma_{\rm cm}	= \frac{E_\chi+E_\ell}{\sqrt{s}}, \qquad V_{\rm cm} = \sqrt{1-\Gamma_{\rm cm}^{-2}} .
	\eeq
	The final lepton energy in the local medium frame can be obtained by a Lorentz boost, and the result is
	\beq
	E'_\ell(\varphi_\ast) =	\mathcal{A}_\ell+ \mathcal{B}_\ell\cos\varphi_\ast, \label{eq:Elprime_ab_app}
	\eeq
	where
	\beq
	\mathcal{A}_\ell =	\Gamma_{\rm cm}	\left[E_\ell^\ast +	V_{\rm cm}p_\ast \cos\eta\cos\theta_\ast \right],\qquad	\mathcal{B}_\ell=\Gamma_{\rm cm}V_{\rm cm}p_\ast \sin\eta\sin\theta_\ast,
	\eeq
	and $\eta$ is the angle between the initial lepton momentum and the boost direction (in CM frame), which is fixed by
	\beq
	\cos\eta =	\frac{E_\ell/\Gamma_{\rm cm}-E_\ell^\ast}{V_{\rm cm}p_\ast}.
	\label{eq:eta_app}
	\eeq
	The CM scattering angle $\theta_\ast$ is given by
	\beq
	\cos\theta_\ast = 1+\frac{2st}{\gamma^2(s)}. \label{eq:theta_star_app}
	\eeq
	
	Finally the zero-temperature azimuthally averaged Pauli-blocking factor can be written as
	\beq
	\overline{\mathcal P}_\ell(E_\ell,s,t;r)	= \left\{
	\begin{array}{ll}1,& \mu_\ell(r)\le {\cal A}_\ell-|{\cal B}_\ell|, \\
		0, & \mu_\ell(r)\ge {\cal A}_\ell+|{\cal B}_\ell| ,\\
		\frac{1}{\pi}	\arccos \left[	\frac{\mu_\ell(r)-{\cal A}_\ell}{|{\cal B}_\ell| }\right],&{\cal A}_\ell-|{\cal B}_\ell| < \mu_\ell(r) <{\cal A}_\ell+|{\cal B}_\ell| .
	\end{array}\right.	\label{eq:Pbar_zeroT_app}
	\eeq
	For ${\cal B}_\ell=0$, Eq.~\eqref{eq:Pbar_zeroT_app} should be understood as $\overline{\mathcal P}_\ell = \Theta\left({\cal A}_\ell-\mu_\ell(r)\right).$
	
    }

	\section{RGEs of model B and validity of the cancellation mechanism}\label{RGEs}
\begin{figure}
		\centering
		\includegraphics[width=0.48\textwidth]{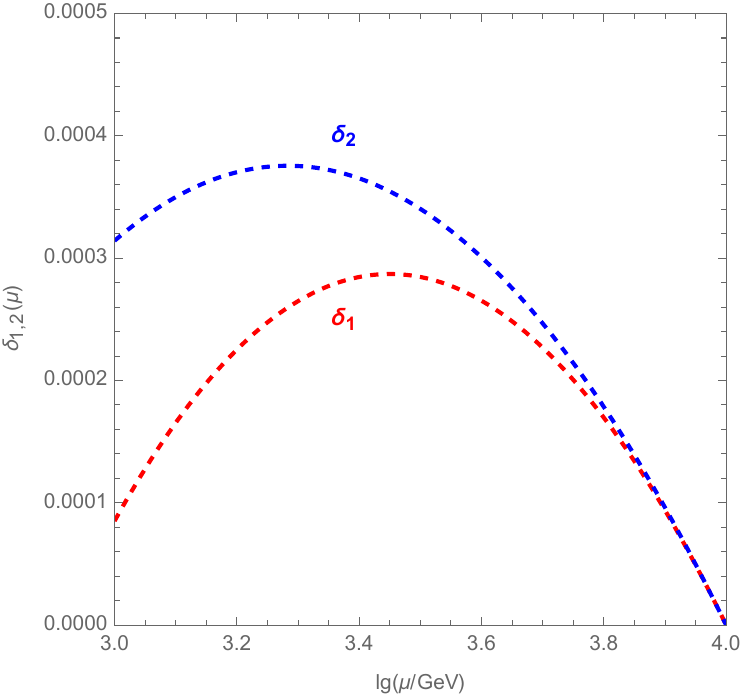}
		\includegraphics[width=0.48\textwidth]{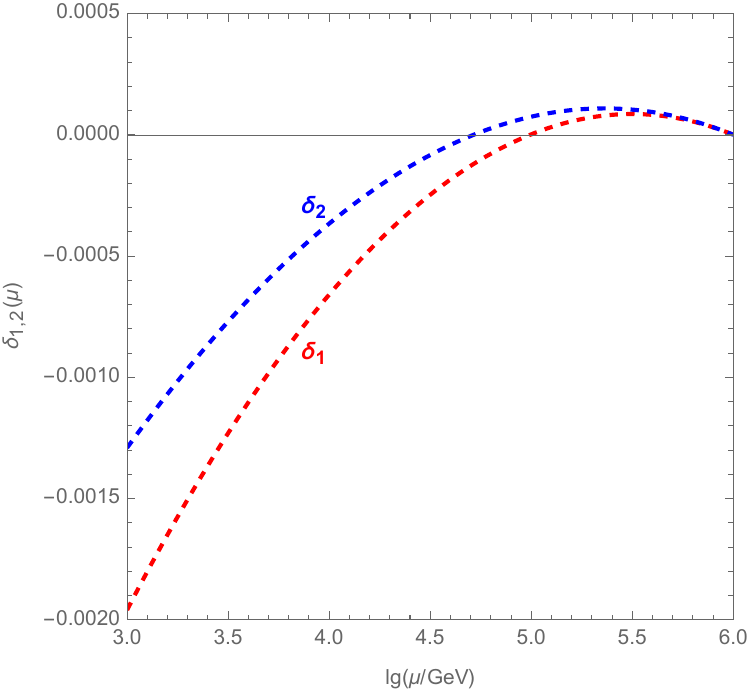}
		\caption{The running of $\delta_{1,2}$ defined in Eq.\eqref{delta12ofmu} for the benchmark \eqref{RGEbm} with $\Lambda_{UV}=10^4~\mathrm{GeV}$ (left panel) and $10^6$~GeV (right panel). }\label{delta12running}
	\end{figure}

	As we discussed in Section~\ref{subsect:pNGBB}, the reliability of cancellation is guaranteed by the relation 
	\beq
	\frac{\lambda_{3\chi}(\mu)}{\lambda_{33}(\mu)}=\frac{\lambda_{H\chi}(\mu)}{\lambda_{H3}(\mu)}\label{can_rel}
	\eeq 
	at the one-loop level. Another way to check this is to solve the following one-loop RGEs of the couplings~\footnote{The RGEs of $\lambda_i$ can be adopted from Ref.~\cite{SalimAdam:2025wlp} by turning off the couplings involving $\eta$ field and Yukawa coupling $\lambda_D$. The gauge coupling contributions to $\beta_{\lambda_i}$s can be adopted from Ref.~\cite{Basso:2010jm}.}:
	\beq
	\left(16 \pi^2\right) \beta_{g_X}&=&\left(4+\frac{1}{3} Q_{\mathrm{DM}}^2\right) g_X^3\label{beta_gX}\\
	\left(16 \pi^2\right) \beta_{\lambda_H} & =&16\pi^2\beta_{\lambda_H}^{(\mathrm{SM})}+2 \lambda_{H 3}^2+4 \lambda_{H \chi}^2, \\
	\left(16 \pi^2\right) \beta_{\lambda_{\chi}} & =&20 \lambda_{\chi}^2+2 \lambda_{3 \chi}^2+8 \lambda_{H \chi}^2-12 Q_{\mathrm{DM}}^2 g_X^2 \lambda_{\chi}+6 Q_{\mathrm{DM}}^4 g_X^4, \\
	\left(16 \pi^2\right) \beta_{\lambda_{33}} & =&18 \lambda_{33}^2+4 \lambda_{3 \chi}^2+8 \lambda_{H 3}^2, \label{beta_l33}\\
	\left(16 \pi^2\right) \beta_{\lambda_{3 \chi}} & =&8 \lambda_{H 3} \lambda_{H \chi}+\lambda_{3 \chi}\left(6 \lambda_{33}+8 \lambda_{\chi}+8 \lambda_{3 \chi}-6 Q_{\mathrm{DM}}^2 g_X^2\right), \label{beta_l3c}\\
	\left(16 \pi^2\right) \beta_{\lambda_{H 3}} & =&\lambda_{H 3}\left(A_{\mathrm{SM}}+6 \lambda_{33}+8 \lambda_{H 3}\right)+4 \lambda_{H \chi} \lambda_{3 \chi}, \label{beta_lh3}\\
	\left(16 \pi^2\right) \beta_{\lambda_{H \chi}} & =&\lambda_{H \chi}\left(A_{\mathrm{SM}}+8 \lambda_{\chi}+8 \lambda_{H \chi}-6 Q_{\mathrm{DM}}^2 g_X^2\right)+2 \lambda_{H 3} \lambda_{3 \chi},\label{beta_lhc}
	\eeq
	where $A_{\mathrm{SM}} \equiv 12 \lambda_H+6 y_t^2-\frac{9}{2} g_2^2-\frac{3}{2} g_Y^2$, and the SM beta function $16\pi^2\beta_{\lambda_H}^{(\mathrm{SM})}$ can be found in any QFT textbook (or articles such as Ref.~\cite{Cai:2015kpa} and many others). The running of gauge coupling Eq.~\eqref{beta_gX} can be solved by itself, while the others are entangled with each other, so usually they can be solved only numerically. To analytically understand why the relation in Eq.~\eqref{can_rel} holds, we evaluate the running of the ratios $\lambda_{3\chi}/\lambda_{33}$ and $\lambda_{H\chi}/\lambda_{H3}$. Using the logarithmic derivative identity $\frac{d}{d\ln\mu} \ln(\lambda_i/\lambda_j) = \frac{\beta_{\lambda_i}}{\lambda_i} - \frac{\beta_{\lambda_j}}{\lambda_j}$, we divide the $\beta$ functions in Eqs.~\eqref{beta_l33}--\eqref{beta_lhc} by their corresponding couplings and subtract the respective equations. This yields:
	\beq
	\left(16 \pi^2\right)  \frac{d}{d\ln\mu}\ln\frac{\lambda_{3\chi}}{\lambda_{33}}  & =&8 \lambda_{H 3} \left[\frac{\lambda_{H \chi}}{\lambda_{3\chi}}-\frac{\lambda_{H 3}}{\lambda_{33}}\right]+\left[8 \lambda_{\chi}+8 \lambda_{3 \chi}-12 \lambda_{33}-\frac{4 \lambda_{3 \chi}^2}{\lambda_{33}}\right]-6 Q_{\mathrm{DM}}^2 g_X^2\nonumber\\
	&=&-6Q_{DM}^2g_X^2+\mathcal{O}\left(\lambda_i\frac{Q_{DM}^2g_X^2}{16\pi^2}\ln\frac{\Lambda_{UV}}{\mu}\right),\label{beta_sub_1}\\
	\left(16 \pi^2\right)  \frac{d}{d\ln\mu} \ln\frac{\lambda_{H\chi}}{\lambda_{H3}} & =& \left[8 \lambda_{\chi}-6 \lambda_{33}+\frac{2 \lambda_{H 3} \lambda_{3 \chi}}{\lambda_{H\chi}}-\frac{4 \lambda_{H \chi} \lambda_{3 \chi}}{\lambda_{H3}}\right]+8\left[ \lambda_{H \chi}- \lambda_{H 3}\right]-6 Q_{\mathrm{DM}}^2 g_X^2,\nonumber\\
	&=&-6Q_{DM}^2g_X^2+\mathcal{O}\left(\lambda_i\frac{Q_{DM}^2g_X^2}{16\pi^2}\ln\frac{\Lambda_{UV}}{\mu}\right).\label{beta_sub_2}
	\eeq
	where we have used the SO(3)-restoration assumption, Eq.~\eqref{eq:SO3_UV_BC_landau_main}, and both second terms of Eqs.~\eqref{beta_sub_1} and \eqref{beta_sub_2} are subdominant if $[\lambda_i\ln(\Lambda_{UV}/m_X)/16\pi^2]\ll 1$. Therefore, we can approximately solve them as
	\beq
	\frac{\lambda_{3\chi}(\mu)}{\lambda_{33}(\mu)}\approx \frac{\lambda_{H\chi}(\mu)}{\lambda_{H3}(\mu)}\approx  \exp\left[\frac{3Q_{\mathrm{DM}}^2 g_X^2}{16\pi^2}\ln\frac{\Lambda_{UV}^2}{\mu^2}\right]\approx \frac{1}{1-\frac{3Q_{\mathrm{DM}}^2 g_X^2}{16\pi^2}\ln\frac{\Lambda_{UV}^2}{\mu^2}}\label{approsol}
	\eeq 
	which is expected, and the rescaling factor also matches the estimation given by Eq.~\eqref{rescale}. In Fig.~\ref{delta12running}, we present the numerical results of
	\beq
	\delta_{1}\equiv \frac{\frac{\lambda_{3\chi}}{\lambda_{33}}-\frac{1}{1-\frac{3Q_{\mathrm{DM}}^2 g_X^2}{16\pi^2}\ln\frac{\Lambda_{UV}^2}{\mu^2}}}{\frac{1}{1-\frac{3Q_{\mathrm{DM}}^2 g_X^2}{16\pi^2}\ln\frac{\Lambda_{UV}^2}{\mu^2}}},\quad \delta_{2}\equiv \frac{\frac{\lambda_{H\chi}}{\lambda_{H3}}-\frac{1}{1-\frac{3Q_{\mathrm{DM}}^2 g_X^2}{16\pi^2}\ln\frac{\Lambda_{UV}^2}{\mu^2}}}{\frac{1}{1-\frac{3Q_{\mathrm{DM}}^2 g_X^2}{16\pi^2}\ln\frac{\Lambda_{UV}^2}{\mu^2}}},\label{delta12ofmu} 
	\eeq
	for a benchmark 
	\beq
	&&\lambda_{33}(\Lambda_{UV})=\lambda_{3\chi}(\Lambda_{UV})=\lambda_{\chi}(\Lambda_{UV})=0.15,~\lambda_{H\chi}(\Lambda_{UV})=\lambda_{H3}(\Lambda_{UV})=0.09,\nonumber\\
	&&m_{h_2}=m_{\mathrm{DM}}=0.6~\mathrm{TeV},~m_X=1.2~\mathrm{TeV},~Q_{\mathrm{DM}}=3.\label{RGEbm}
	\eeq
	$\delta_{1}~(\delta_{2})$ represent the relative errors between $\frac{\lambda_{3\chi}(\mu)}{\lambda_{33}(\mu)}$ $\left(\frac{\lambda_{H\chi}(\mu)}{\lambda_{H3}(\mu)}\right)$ and $r\equiv\frac{1}{1-\frac{3Q_{\mathrm{DM}}^2 g_X^2}{16\pi^2}\ln\frac{\Lambda_{UV}^2}{\mu^2}}$. We can see that these errors are tiny, and thus our approximated formula Eq.~\eqref{approsol} works very well.

\begin{figure}
		\centering
		\includegraphics[width=0.48\textwidth]{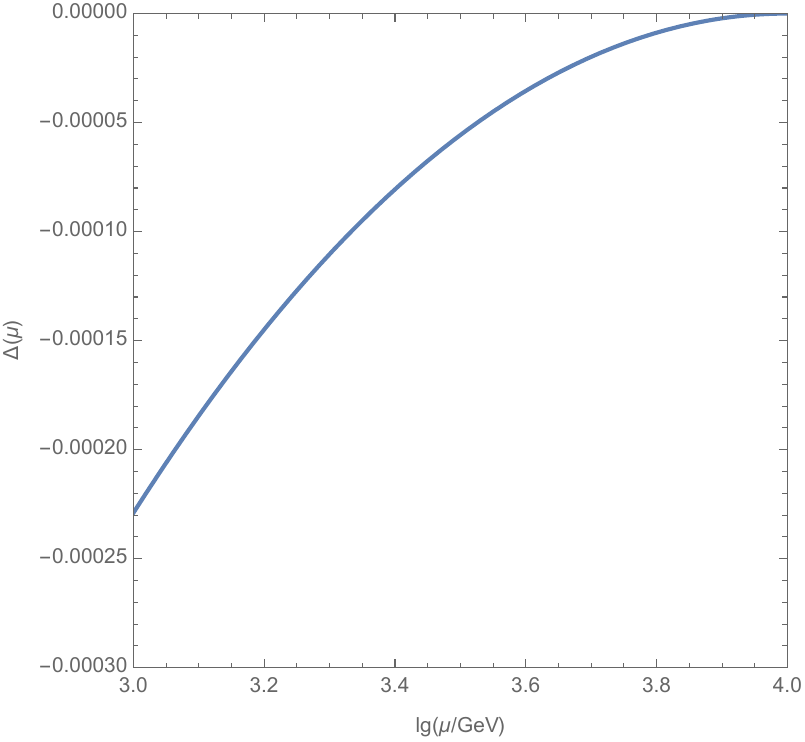}
		\includegraphics[width=0.48\textwidth]{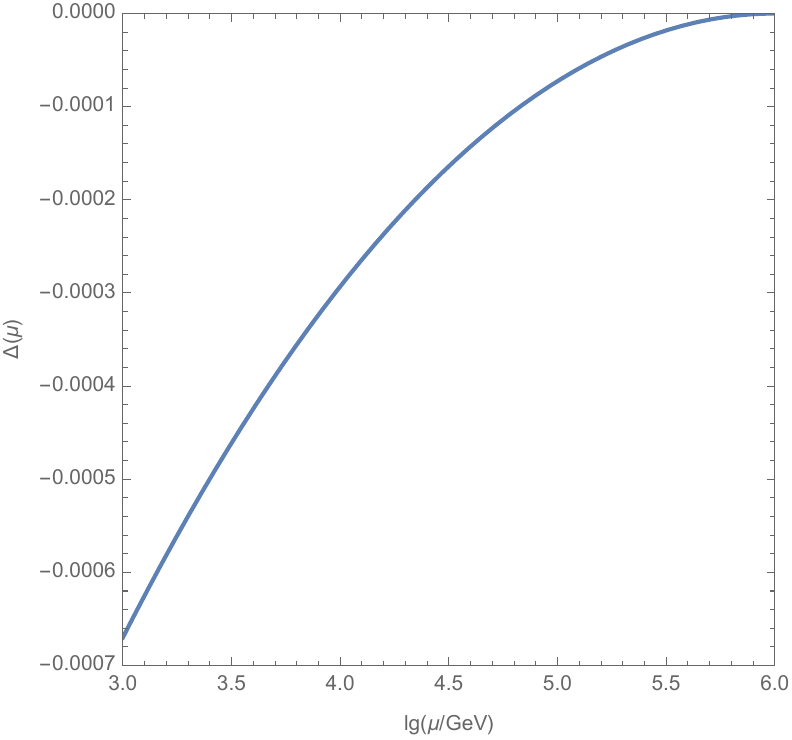}
		\caption{The running of $\Delta$ defined in Eq.\eqref{Deltaofmu} for the benchmark \eqref{RGEbm} with $\Lambda_{UV}=10^4~\mathrm{GeV}$ (left panel) and $10^6$~GeV (right panel). }\label{Deltarunning}
	\end{figure}

	We also show the running of the relative error
	\beq
	\Delta(\mu)\equiv \frac{\lambda_{3\chi}\lambda_{H3}-\lambda_{33}\lambda_{H\chi}}{\lambda_{3\chi}\lambda_{H3}}\label{Deltaofmu}
	\eeq
	in Fig.~\ref{Deltarunning} for the same benchmark. We can see that the errors are quite tiny, implying that the cancellation is robust in this model.
	
	\bibliographystyle{apsrev4-1-JHEPfix}
	\bibliography{ref}
\end{document}